\documentclass[twocolumn,usegraphicx,usenatbib,useAMS]{mn2e}

\newcommand{\bit}{\begin{itemize}}
\newcommand{\eit}{\end{itemize}}
\newcommand{\nit}{\noindent}

\def\H{\hbox{H}}
\def\HI{\hbox{H$\scriptstyle\rm I $}}

\def\C{\hbox{C}}
\def\CII{\hbox{C$\scriptstyle\rm II $}}
\def\CIII{\hbox{C$\scriptstyle\rm III $}}
\def\CIV{\hbox{C$\scriptstyle\rm IV $}}
\def\Si{\hbox{Si}}
\def\SiII{\hbox{Si$\scriptstyle\rm II $}}
\def\SiIII{\hbox{Si$\scriptstyle\rm III $}}
\def\SiIV{\hbox{Si$\scriptstyle\rm IV $}}
\def\O{\hbox{O}}
\def\OI{\hbox{O$\scriptstyle\rm I $}}
\def\OVI{\hbox{O$\scriptstyle\rm VI $}}
\def\OVII{\hbox{O$\scriptstyle\rm VII $}}
\def\OVIII{\hbox{O$\scriptstyle\rm VIII $}}
\def\MgII{\hbox{Mg$\scriptstyle\rm II $}}

\def\Lya{Ly$\alpha$~}

\newcommand{\beq}{\begin{equation}}
\newcommand{\eeq}{\end{equation}}
\newcommand{\ba}{\begin{eqnarray}}
\newcommand{\ea}{\end{eqnarray}}
\newcommand{\etal}{et al.\ }

\def\gtsima{$\; \buildrel > \over \sim \;$}
\def\ltsima{$\; \buildrel < \over \sim \;$}
\def\gsim{\lower.5ex\hbox{\gtsima}}
\def\lsim{\lower.5ex\hbox{\ltsima}}
\def\msun{{M_\odot}}
\def\zsun{{Z_\odot}}

\begin{document}
\title[Absorption features of high redshift galactic winds]{Absorption features of high redshift galactic winds}
 \author[A.~P.~M.~Fangano et al.]{A.~P.~M.~Fangano$^1$\thanks{E-mail: fangano@astro.uni-bonn.de},
A.~Ferrara$^2$ \& P.~Richter$^{1,3}$\thanks{DFG Emmy-Noether Fellow}\\
$^1$ Argelander-Institut f\"ur Astronomie\thanks{Founded by merging of the Institut f\"ur Astrophysik und Extraterrestrische Forschung, the Sternwarte and the Radioastronomisches Intitut der Universit\"at Bonn.}, Universit\"at Bonn, Auf dem H\"ugel 71, 53121 Bonn, Germany \\
$^2$ SISSA/International School for Advanced Studies, Via Beirut 4, 34014 Trieste, Italy \\
$^3$ Institut f\"ur Physik, Universit\"at Potsdam, Am Neuen Palais 10, 14469 Postdam, Germany}

\maketitle

\begin{abstract}

\nit The environment of high-redshift galaxies is characterized by
both wind-driven outflowing gas and gravitationally infalling
streams. To investigate such galaxy-IGM interplay we have generated
synthetic optical absorption line spectra piercing the volume
surrounding a starbursting analog of a Lyman Break Galaxy selected
in a $z \approx 3$ output from a SPH simulation, including a
detailed treatment of mechanical feedback from winds. Distributions
for several observable species (\HI, \CIII, \CIV, \SiII, \SiIII,
\SiIV, \OVI, \OVII, and \OVIII ) have been derived by
post-processing the simulation outputs. The hot wind material is
characterized by the presence of high-ionization species such as
\OVI, \OVII, and \OVIII$~$(the latter two observable only in X-ray
bands); the colder ($T<10^{5.5}$ K) infalling streams can be instead
identified by the combined presence of \SiII, \SiIII, and
\CIII$~$optical absorption together with \OVI$~$that surrounds the
cooler gas clumps. However, both line profile and Pixel Optical
Depth analysis of the synthetic spectra show that the intergalactic
filament in which the wind-blowing galaxy is embedded produces
absorption signatures that closely mimic those of the wind
environment. We conclude that it may be difficult to clearly
identify wind-blowing galaxies and their complex gaseous environment
at high redshift in optical QSO absorption-line spectra based solely
on the observed ion absorption patterns.

\end{abstract}


\section{Introduction}

Shortly after the first detection of quasi-stellar objects (quasars
or QSOs), the presence of many narrow absorption lines in the
spectra of these objects was recognized for the first time (e.g.,
\citet{Bahcall66}). It soon became clear that these lines
(``Ly$\alpha$ forest") are produced by absorption of neutral
hydrogen in gaseous structures that fill the intergalactic space
(the InterGalactic Medium, IGM). Today it is widely accepted that
the neutral hydrogen Ly$\alpha$ forest traces baryon density
fluctuations associated with gravitational instability as part of
the large-scale structure formation in the Universe.

With the advent of more advanced spectrographs and larger
telescopes, higher quality spectra were obtained and with great
surprise it was discovered that the hydrogen forest is complemented
by absorption lines from heavy elements \citep{Cowie95}. Subsequent
analysis (\citet{Ellison99}, \citet{Schaye00}, \citet{Songaila01},
\citet{Simcoe06}) have shown that the IGM is enriched with metals at
virtually any overdensity and redshift probed, as shown by the
observation of resonance transitions associated to ions like \CIII,
\CIV, \SiII, \SiIII, \SiIV, \OVI, and others; the enrichment of
underdense regions is still matter of debate \citep{Aracil04}. A
major problem in our current understanding of the IGM is to
determine when and by what transport means metals have been able to
travel to large distances from their productions sites (presumably,
stars in galaxies.  It is then clear that the answers to these
questions have to come from the study of the physical galaxy-IGM
interplay.

One possible mechanism is dynamical stripping. During a close
encounter of two or more galaxies, tidal forces can strip away from
them part of the enriched gas, which is then dispersed into the IGM.
This scenario is particularly important at high redshift, when the
encounter rate is enhanced. However, this mechanism turns out to be
too inefficient in reproducing the observed metal distribution with
H\,{\sc i} column densities \citep{Aguirre01}. A second mechanism is
based on the idea that the energy/momentum deposited by multiple
supernovae can drive an outflow of material from the host galaxy
(\citet{Dekel86}; \citet{MacLow99}, \citet{Ferrara00};
\citet{Madau01}; \citet{Ferrara05}; \citet{Springel03};
\citet{Aguirre05}; \citet{Dave06}). In this scenario, multiple
supernova-driven bubbles merge together to form a "superbubble"; the
entrapped material would blow-out from the galaxy and vent the hot
interior (largely enriched by freshly produced heavy elements) into
the IGM giving rise to a large-scale outflow. Yet, the epoch when
most of enrichment took place is still unclear. Studies carried out
by \citet{Songaila01}, \citet{Schaye03}, \citet{Songaila05},
\citet{Ryan-Weber06} have shown that the observed \CIV$~$absorption
characteristics do not evolve significantly in the redshift range $2
< z < 6$.  This is surprising in view of the fact that most of the
star-formation activity in the universe took place at $z < 5$ (see
\cite{Hopkins04} for an updated compilation). The findings were
interpreted as the results of an early ($z>6$) metal pollution
(often referred to as {\it pre-enrichment}) by the first generation
(Pop~III) of stars and galaxies (\citet{Madau01},
\citet{Scannapieco02}).

With the combined analysis of QSO and galaxy spectra it is possible
to connect the properties of the intergalactic absorption lines with
that of the foreground galaxies. Using this technique
\cite{Adelberger03} and \cite{Adelberger05} demonstrated that almost
all galaxies at $z>2$ can be associated with metal absorbers, in
agreement with the earlier theoretical study by \citet{Haehnelt96},
and subsequent analysis by \citet{Desjacques04}, \citet{Maselli04},
\citet{Bertone05}, and \citet{Kawata07}. While this points directly
to the hypothesis that those galaxies are experiencing large scale
outflows and hence polluting the IGM {\it in situ}, the actual mass
loss rate and, most importantly, whether they are able to escape the
gravitational pull of the galaxy are very hard to assess.
\citet{Simcoe06_1} have recently analyzed a line of-sight toward a
quasar at $z\simeq$ 2.7 and searched for a correlation between
absorbers and nearby galaxies. By constructing detailed ionization
models for the different absorbers, they find that their
observations indicate a metal-enriched volume of $\sim 100-200
h_{71}^{-1}$ physical kpc around the analyzed galaxies. Furthermore,
the sheet-like geometry of the absorbing clouds ($\Delta L < 1$ kpc)
as well as their high metallicity ($Z > 0.1-0.3 Z_{\odot}$) suggest
an outflow scenario for their origin. Concentrating on low
redshifts, ($z\simeq$ 0.5) \citet{Bouche06}, starting from a sample
of 1806 \MgII$~$absorbers, show that their equivalent widths ($W >
2$~\AA) are anticorrelated with the absorber halo-mass. Since this
indicates that absorbing clouds are not in virial equilibrium with
the host-galaxy, they concluded that these absorbers are most
probably generated by a starburst activity and expelled via
supernova-driven winds. Note that an early-epoch enrichment model is
not excluded by this scenario since metals ejected by the first
galaxies ($z \simeq$ 6-12) should also end up in overdense regions
at lower redshifts (\cite{Porciani05}, \cite{Scannapieco05}).

The complexity of absorbing cloud populations asks for a careful
analysis and interpretation of observational data. For this aim,
simulations can be quite helpful, especially in trying to identify
the various gas components that contribute to the gas flow in the
intergalactic environment of galaxies. However, simulations are not
yet able to follow in detail the full range of physical phenomena
that characterize star formation and feedback. So far, the coupling
of a consistent modelling of the ISM with a reliable prediction of
the physical properties of galactic winds turns out to be quite
difficult. New generation numerical codes include often an explicit
recipe for starting galactic winds in suitable situations. In these
cases of {\it defaillance}, simulations have to rely on
phenomenological prescriptions deduced from theoretical models or
inferred from the analysis of the local Universe. To date, the
number of observed galaxies with on-going outflows is quite
consistent with the theoretical predictions, but the correlation
between the outflow and galaxy properties is difficult to be
interpreted and shows no substantial trends (\citet{Martin99},
\citet{Heckman00}). Recently, observations made by \cite{Martin05}
and \cite{Murray05} showed a good agreement with theoretical models
for momentum-driven winds. In this model, radiation impinges on dust
grains, transferring this way its momentum. The latter is moving out
of the galaxies and  then may couple with the surrounding gas and
start an outflow \citep{Murray05}. Based on this model,
\cite{Oppenheimer06} have run a set of numerical simulations using
the code GADGET2 \citep{Springel05} with a span of feedback
prescriptions for galactic winds. They demonstrated that
momentum-driven outflows, calibrated via the local universe
observations, are able to reproduce a wide variety of \CIV~absorbers
statistics as observed in QSOs spectra.

In this framework of these recent studies we have performed our
study. We have used an output from a GADGET2 cosmological simulation
that uses state-of-art treatment of multiphase ISM and winds, and
keeps track of the contribution of Pop~III stars to the overall
metal budget. These simulations have been then analyzed with the
intention of providing quantitative and clear-cut diagnostics for
observational investigation of outflow phenomena around high
redshift galaxies; to this aim, and also to clarify the association
between absorbing lines and galactic environments, we have performed
a spectroscopic analysis around a prototypical galaxy characterized
by ongoing wind-activity.

This paper is organized as follows: in \S2, we briefly describe the
numerical simulation; we then move to the galaxy selection criteria
and to the analysis of its gaseous environments. In \S3 we analyze
the metal absorption features of the "wind region" observed in our
spectra and we discuss briefly of the \Lya forest around our target
galaxy. In Section \S4 we give then our conclusions. In addition, we
provide in Appendix A details about the method used to generate our
spectra and a summary about the considered metal transitions.
Finally, in Appendix B we show our synthesized spectra.

\section{Numerical Simulations}

Our analysis is based on a hydrodynamic cosmological simulation. The
simulation adopts a $\Lambda$CDM cosmological model with
$\Omega_{M}$=0.3, $\Omega_{\Lambda}$=0.7, $\Omega_{b}$=0.04,
$h=0.7$, and $\sigma_{8}=0.8$; periodic boundary conditions are
adopted. The numerical code used for the simulation is a modified
version of GADGET \citep{gadget} modified according to
\citet{Marri03}, where the interested reader can find the details of
the new algorithms introduced. In brief, additions concern: (i) an
improved treatment of the multiphase nature of the gas, and (ii) a
more refined recipe for stellar feedback. The ISM is modeled as a
two-phase medium, made of a cold, dense component and a warm,
diffuse one. Once star formation occurs, the feedback energy is
distributed to the surrounding particles with a fraction $e_c$
($e_w$) incorporated in the cold (warm) phase. Values for $e_c$ and
$e_w$ could be fixed from a complete theory of interstellar medium,
or by specific numerical simulations; in our case these constants
were both fixed to 0.4. The remaining 20\% of the energy is assumed
to be radiated away.

The scheme also tracks the gas metal enrichment due to supernova
(SN) explosions by considering a heavy element yield $y_Z$ such
that: \beq \delta M_Z= y_Z \delta M_\star\, \label{metal_prod} \eeq
where $\delta M_Z$ is the total metal mass produced per stellar mass
($\delta M_*$) formed. Once metals are produced, a fixed fraction
$f_Z$=0.3 is distributed among the same hot neighbors which receive
the feedback energy. Tracking the metallicity allows us to apply the
so-called `critical metallicity' criterion introduced by Schneider
\etal (2002). According to this criterion, the Initial Mass Function
(IMF) of star forming sites is biased towards very massive (PopIII)
stars if the metallicity is below $Z_{cr}=10^{-4} \zsun$; above this
threshold a switch to a standard stellar population (PopII) with a
Salpeter IMF occurs. These stellar populations have distinct
feedback prescriptions, depending on the explosion energies,
$e_{sn}$, and metal production efficiencies, $y_Z$ of PopIII and
PopII stars. Therefore, we adopt the following set of parameters:

\beq e_{sn}^{III}=2.3 \times 10^{50} {\rm ~~erg ~\msun^{-1}}\;
~~~y_Z^{III}=0.637\; \label{popII_param} \eeq for PopIII stars, and
\beq e_{sn}^{II}=4.0 \times 10^{48} {\rm ~~erg~\msun^{-1}}\;
~~~y_Z^{II}=0.004\; \label{popII_param} \eeq for PopII stars. These
above parameters are not yet fully fixed by theory, as the outcome
of stellar explosions depend e.g. on progenitor composition, nuclear
reactions rates, explosion energy.

We have run simulations at two different resolutions. The low
resolution simulation, used mostly as a control run, consists of
64$^3$ particles simulating a 7.0 $h^{-1}$ comoving Mpc; the high
resolution run is made up of 128$^3$ particles in a 10.5 $h^{-1}$
comoving Mpc. Both have a mass resolution of $M\approx 10^6 \msun
h^{-1}$. In both runs the evolution was followed including the
effect of a UV-background (produced by QSOs and galaxies and
subsequently filtered through the IGM) whose shape and amplitude
were taken from \citet{UV_bck}. Unless otherwise specified we report
results based on the high resolution run.

\subsection{Target galaxy}

\begin{table}
\begin{center}
\begin{tabular}{cccll}
\hline
\multicolumn{1}{c}{Element}&\multicolumn{1}{c}{Tracer}&\vline&\multicolumn{2}{c}{Abundances}\\
\hline \hline
$[C/H]$&$\CIV$&\vline&$-2.58$&$-2.82^1, -3.47^2$\\
$[O/H]$&$\OVI$&\vline&$-2.14$&$-2.82^1, [-2.22, -1.3]^3$\\
$[Si/H]$&$\SiIV$&\vline&$-1.99$&$-2.00^4$\\
\hline
\\
\multicolumn{5}{l}{$^1$ \citet{Schaye03}, $^2$ \citet{Simcoe04}}\\
\multicolumn{5}{l}{$^3$ \citet{Telfer02}, $^4$ \citet{Aguirre04}}\\
\end{tabular}
 \caption{Abundances values relative to solar from our low resolution run:
shown are the considered element (first column), transition used as
tracer (second), median value retrieved from our simulation (third),
observed values (forth). For a meaningful comparison with
observations, we have included in our analysis only pixels with \Lya
and metal optical depth higher than $10^{-3}$.} \label{tab_metal}
\end{center}
\end{table}

\begin{figure}
\includegraphics[width=9cm]{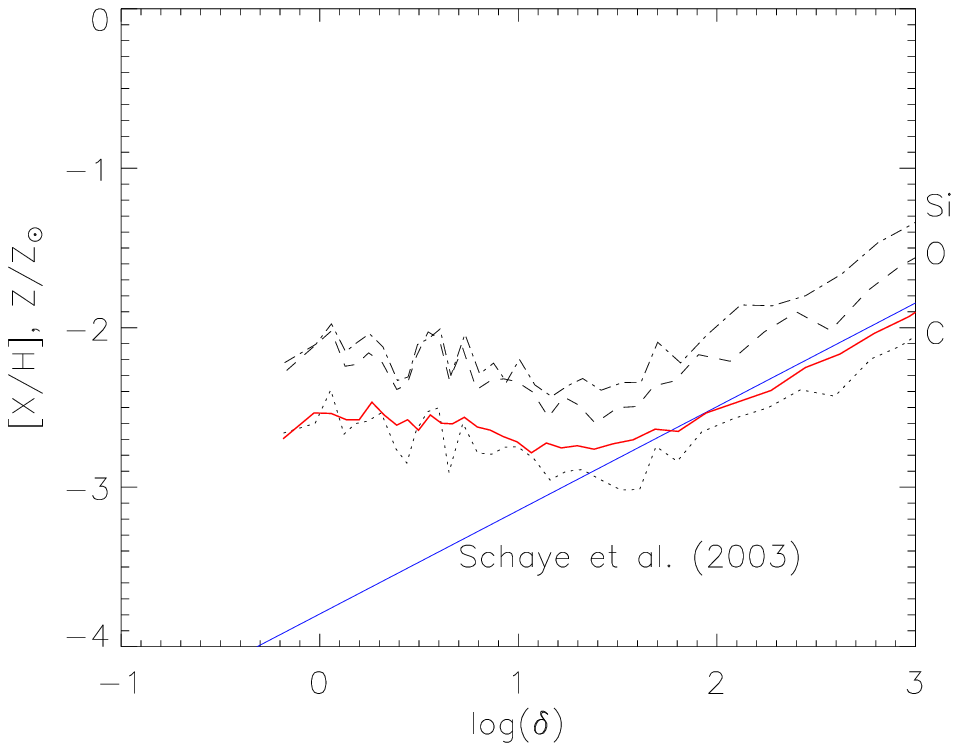}
\caption{Abundances relative to solar for \C$~$(dotted),
\O$~$(dashed) and \Si$~$(dot-dashed) as a function of overdensity.
The thick red curve represents the values for the total metallicity
relative to solar while the straight blue line show the relation
found by \citet{Schaye03} for [\C/\H]  (eq.\ref{C_H_Schaye}).}
\label{mean_abb}
\end{figure}
As general test for our metal feedback recipe, we have analyzed the
low resolution run for the average metallicity of the IGM as traced
by carbon, oxygen and silicon. In Fig. \ref{mean_abb} we show the
relation of these abundances, relative to solar, as a function of
the overdensity ($\delta \equiv \rho/<\rho>$). In order to compare
with observations, we have limited our analysis to those pixels
characterized by optical depths for both \Lya and the metal
transition $>10^{-3}$. From the plot we deduce that from our
simulation a small raise in the metal abundances should be expected
at low overdensities where wind effects are more evident. The
combined action of our thermal and chemical feedback seems to
suggest also the fact that the [\C/\H] ratio could be the closest
numerically tracer of the total metal content of the gas. In the
Figure we also show the relation found by \citet{Schaye03} for
[\C/\H] vs. overdensity: \beq
[\C/\H]=-3.47^{+0.07}_{-0.06}+0.65^{+0.10}_{-0.14}\times(\log \delta
- 0.5). \label{C_H_Schaye}\eeq Tab. \ref{tab_metal} compares the
simulated mean abundances of the three elements/tracers with the
observational values. From the joint analysis of Fig. \ref{mean_abb}
and Tab. \ref{tab_metal} we conclude that our model is quite
successful in reproducing both Schaye relation and the level of
pollution in the IGM as derived by QSOs absorption lines. This
strengthens our confidence on the implemented feedback scheme.

In order to quantify and diagnose the effects of feedback
and winds on galactic environments, the first step is to identify a
suitable representative target galaxy in the simulation box showing
a strong wind. To this aim, we have searched in the high-resolution
run output at $z=3.26$ for high density ($n > 10^{-2}$~cm$^{-3}$),
high temperature ($T > 10^6$~K) particles. The latter condition
ensures that we are dealing with a galaxy that is surrounded by
shocked gas; however, such high temperatures can also be reached
during virialization and/or infall, so they do not necessarily imply
that the object is experiencing a wind phase. We thus complement the
high-temperature condition with that of a high peculiar particle
velocity, as infalling material is likely slowed down by the
interaction with the halo gas. The total mass of the selected target
galaxy which fulfills the above listed conditions is $\approx
10^{11} \msun$, i.e. a mass typical for Lyman Break Galaxies (LBG).
In the chosen volume, the average metallicity is $\approx 10^{-2}
Z_\odot$.

\subsection{The target galaxy environment}
\begin{figure*}
\includegraphics[width=14cm]{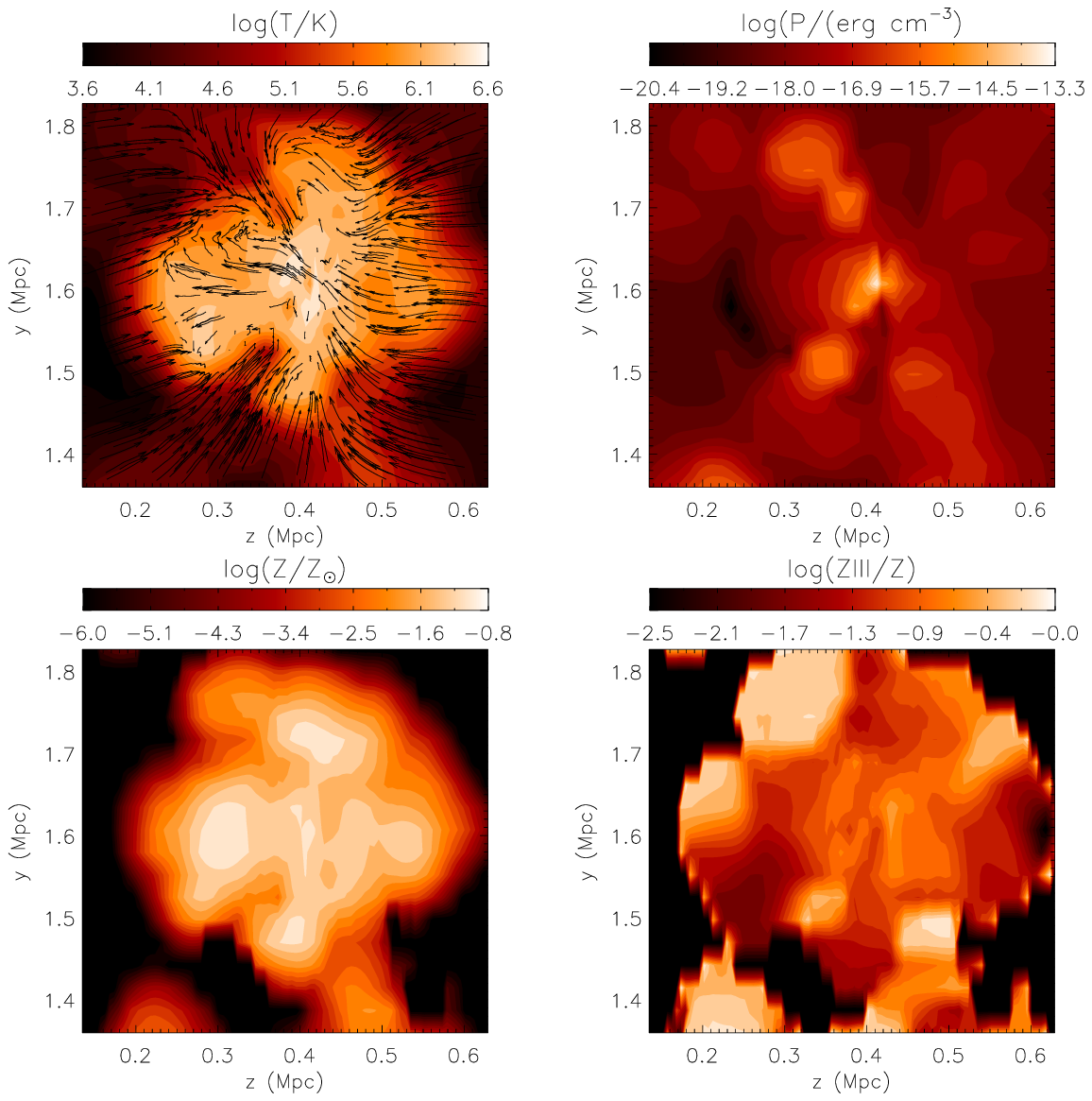}
\caption{Maps of various quantities across the target galaxy center.
{\it Upper left}: temperature and peculiar velocity field (arrows;
the largest one corresponds to 250 km~s$^{-1}$); {\it Upper right}:
ram-pressure; {\it Bottom left}: total metallicity; {\it Bottom
right}: PopIII-to-total metallicity ratio. Sizes are in physical
units.} \label{composit_bubble}
\end{figure*}

In Fig. \ref{composit_bubble} we show 2D maps of different
quantities (temperature and velocity field, ram-pressure, total
metallicity, Pop~III stars contribution to the latter) in a plane
across the target galaxy center. The signature of the wind is
clearly seen as a region of enhanced temperature in the
corresponding map, with a typical (physical) size $\simeq 0.4$ Mpc.
The shock caused by the outflow has heated up the gas to a
temperature $\simeq 10^{6}$ K, but the outflowing gas is being
slowed down by the infalling material. We refer to this region as
the (wind) {\it bubble}.

The interplay between the infalling and outflowing material shapes
the bubble geometry, producing significant deviations from the
commonly assumed spherical symmetry. Instead, we find that the shape
of the bubble is better described by a pancake-like geometry with a
radius-to-thickness ratio of $\simeq 2:1$. The bubble extends almost
orthogonally to the {\it filament} that hosts the galaxy, with 4 or
more lobes caused by the infalling material, which we denote as
(infalling) {\it streams}. An interface between the wind bubble and
a stream is clearly identified by the low pressure arc (right top
panel) to the left of the bubble; there, the outflowing material is
actually stopped by the ram-pressure of the infalling gas causing
the formation of a thin, dense shell. The analogous shell formed at
the top lobe is barely recognizable as it has already started to
recollapse onto the galaxy. To the right hand side of the bubble
instead the collapse evolution is more advanced and the infalling
material is almost freely penetrating into the hot bubble. As a
result, the velocity field is very complex with regions of
outflowing and infalling motions with peculiar velocities up to 250
km s$^{-1}$. Inside the galactic environment, the infalling material
is re-directed by the galactic activity and trapped within the hot
bubble. From the maps shown here we cannot identify any outflow
channel through which the gas could escape the bubble. Inside the
bubble, the material now is heated to temperatures $T \gsim
10^{6}$~K.

The spatial distribution of the metallicity (lower panels) is very
similar to the temperature, clearly indicating that the bubble has
been created by supernova explosions within the target galaxy.
However, the metallicity pattern corresponding purely to PopIII
stars requires a previous star formation activity which pollutes the
streams. This is deduced from PopIII-enriched cold clumps falling
toward the galaxy seen in the upper portion of the map. Such clumps
have been produced at earlier cosmic times by progenitors of the
target galaxy and later became part of the accretion process feeding
it. They are pushed and compressed by the pristine material lying
behind them and streaming from the filament into the bubble (see
temperature map). These kind of condensed clumps in the bubble had
enough time to cool down to temperature $T < T_W \equiv 10^{5.5}$ K;
throughout the following we identify these structures as {\it cold
clumps}. Note that also the filament contains metals, the chemical
composition being dominated by the first generation of stars. We
find that PopIII contribution to the total metal content of the
filament is $> 30 \%$. Metal absorption arising in filamentary gas
therefore is important to study the nucleosynthesis of the first
objects.

In summary, the galaxy environment is characterized by two different
environments: (i) the filament in which the galaxy is embedded, and
(ii) the wind environment, in turn constituted by the wind bubble,
infalling streams and cold clumps accreting onto the galaxy.

\section{Ionization analysis}

Having selected the target galaxy and characterized its environment,
we now investigate their properties by means of synthetic absorption
spectra that can be directly compared with observed QSO
absorption-line data. To this aim, we have analyzed $\sim$ 1600
lines of sight (LOS) through the selected volume\footnote{The
selected region size is of $1.2 \times 0.5 \times 0.5$ physical
Mpc.}.  The ionization equilibrium, under the influence of the UV
background adopted in the simulation, has been calculated in each of
the ($\sim$ 100) cells into which each LOS has been segmented, using
the code CLOUDY96\footnote{http://www.nublado.org}
\citep{Ferland98}. For the following analysis we have used either
the Pixel Optical Depth (POD) technique described in \citet{Cowie98}
and \citet{Aguirre02}, or the automated Voigt-profile fitting using
the code AUTOVP \citep{daveAUTOVP}, depending on the specific
question addressed. For further details on the simulated spectra we
refer to Appendix A.

\begin{figure}
\includegraphics[width=8.5cm]{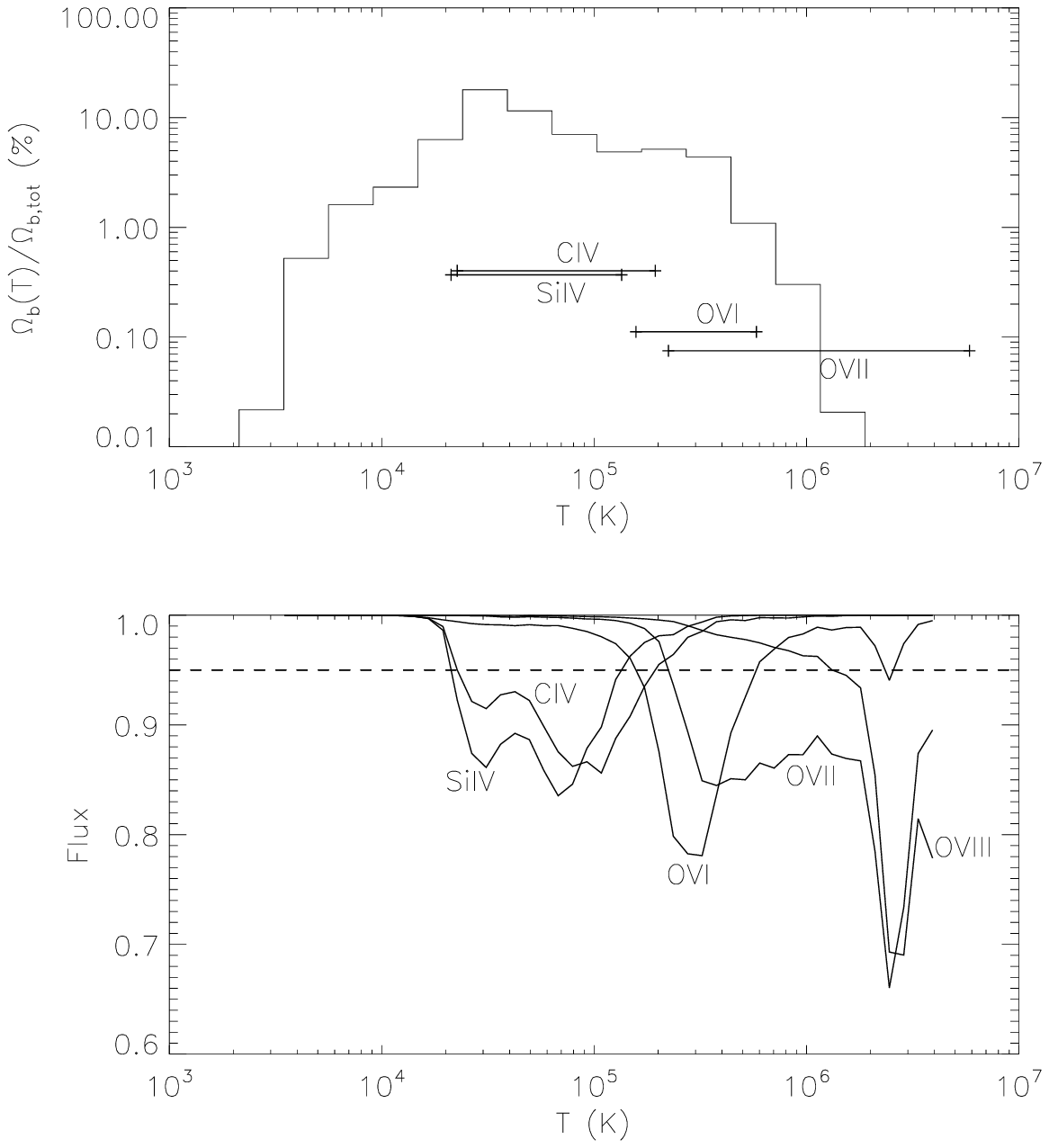}
\caption{{\it Upper panel}: temperature distribution of the baryonic fraction.
Horizontal lines mark the temperature range in which the indicated ion produces a mean
transmission of $\simeq$0.95; their vertical positions refer to the fraction of cosmic
baryons polluted with that ion. {\it Lower panel}: mean transmitted flux, $F=e^{-\langle \tau\rangle}$,
for the different ionic species as a function of gas temperature.} \label{omega_b_T}
\end{figure}
As a first step to search for possible tracers of circumgalactic
high-temperature regions, we have analyzed the gas temperature
distribution shown in Fig. \ref{omega_b_T} (upper panel). The
horizontal width of these lines denotes the temperature range in
which the transmitted flux, $F=e^{-\langle \tau \rangle}$, due to a
given ion opacity is $F< 0.95$, a level denoted also by the dashed
line in the lower panel.  The vertical position of the lines
indicates the fraction of cosmic baryons polluted with that ion (we
have defined a pollution threshold by requiring a ionization
fraction $> 10^{-2.5}$ for each species) integrated over the above
temperature range. The distribution shows a peak at $T \approx
2\times 10^4$~K, accounting for about 20\% of the baryonic fraction
in the volume; an almost equal amount of matter resides at $T \ge
10^5$~K. \CIV$~$traces almost the same amount (about 0.6\%) of
baryons as \SiIV, but it is sensitive to a slightly higher
temperature regime. A similar trend is found for oxygen ions with
\OVI$~$spanning a relatively narrow  range around the peak
temperature ($\approx 3\times 10^5$~K) for collisional ionization of
\OVI, whereas \OVII$~$ extends for almost 1.5 dex. We find that
0.1-0.2\% of all the baryons should be already contaminated with these
species at $z\approx 3.26$.

The mean transmitted flux for these ions varies as a function of
temperature. These profiles are useful to understand the physical
origin of a given ion. The double peak feature seen for \CIV,
\SiIV$~$and \OVII, shows the distinct effect of the radiative and
collisional ionization processes, with the high temperature peak
being associated with collisionally ionization and the other one
(typically less pronounced) resulting from photoionization.
Therefore, for any given species, the relative strength of these two
features gives a measure of the importance of photo- and collisional
ionization. In particular, \CIV$~$has a higher contribution from
collisional ionization than \SiIV. An important point is that both
\OVI$~$and, even more evidently, \OVII$~$are produced by
collisional ionization, and therefore they represent particularly
useful diagnostics to study the hot intergalactic environments like
wind cavities.

In the following we characterize in detail the absorption patterns associated with
the two types of target galaxy environments introduced above, i.e. the wind and the filament environment.

\subsection{Wind environment tracers}

\begin{figure*}
\includegraphics[height=22cm]{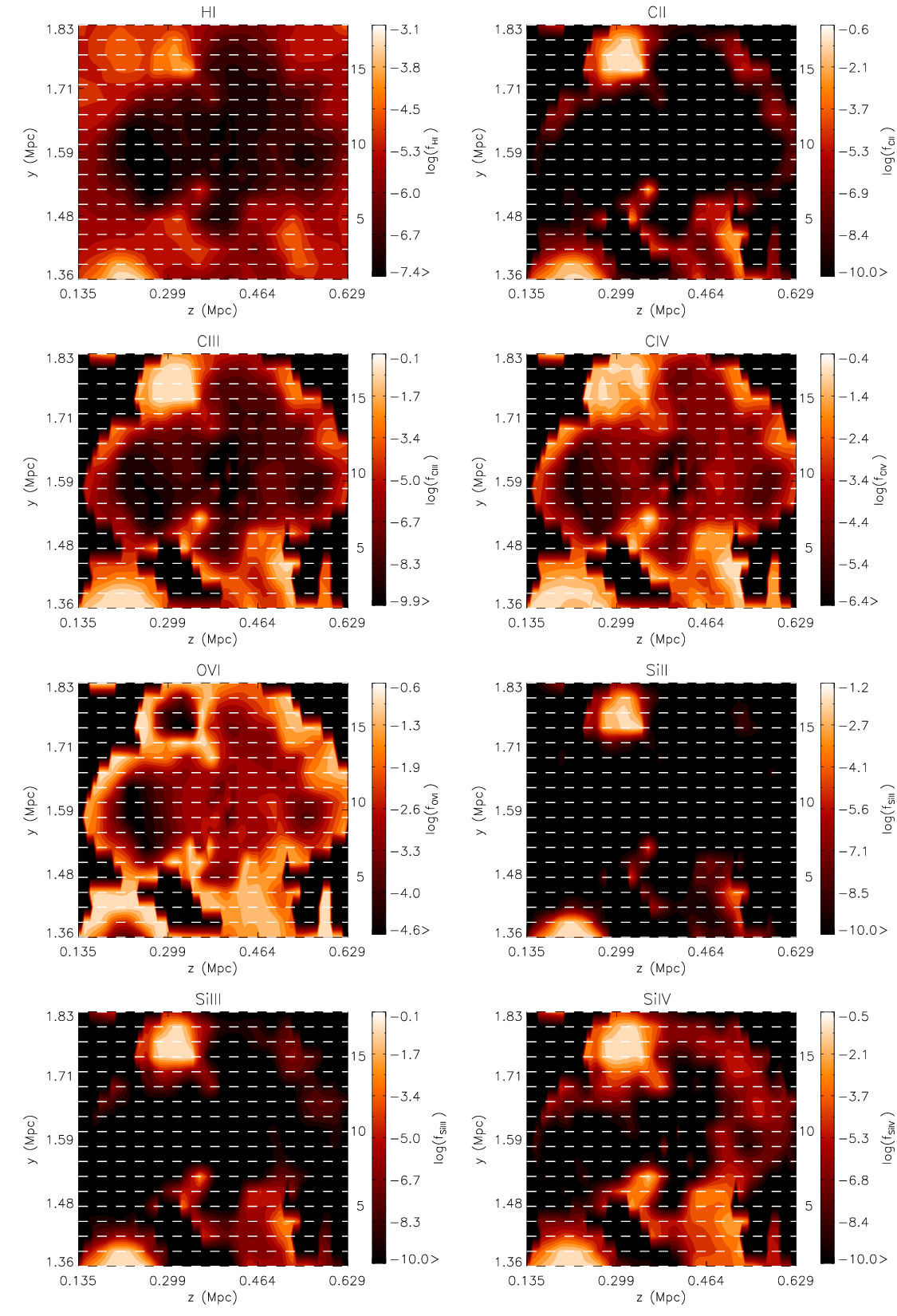}
\caption{Ionization fractions (i.e, the fraction of the $X$ element
in the ionization state $i$ with respect to its total abundance,
$f_{X,i}=n_{X,i}/n_X$) maps across the target galaxy center for
various ions are shown. The dotted lines indicate the analyzed line
of sight (LOS), numbered from 1 to 18 as labeled to the right of
each panel.} \label{frac_ion}
\end{figure*}

As already pointed out, inside the volume enclosed by the
supernova-driven shock three different gas phases/structures
coexist: infalling streams, cold clumps and hot, outflowing gas. As
seen from Fig. \ref{frac_ion}, in which we show the spatial
distribution of the ionization fractions of various ions, the bubble
interior is characterized by a very low \HI$~$abundance ($\log
f_{\rm HI}< -6$), caused by the enhanced collisional ionization rate
associated with the high gas temperature. Such low abundance in the
central regions is common to most of the species shown, with the
only significant exception of \OVI. The latter species appears to be
almost homogeneously distributed in the volume, although its
abundance drops in isolated regions: the most noticeable example are
the \OVI$~$``voids" seen on the upper-left part of the distribution
map for this ion.  Guided by the previous discussion on the nature
of the different structures in the target galaxy environment, we
interpret such \OVI$~$voids as cold clumps that condensed out of the
expelled material and now falling back towards the galaxy.  In
addition to \OVI$~$(which appears to be the best wind tracer) the
bubble region is characterized by considerable abundances of
\OVII$~$and \OVIII$~$(for a detailed view see Appendix C, Fig.
\ref{spectra_bubble_OVII} and Fig. \ref{spectra_bubble_OVIII}).
Although currently unavailable in high-redshift observations these
ions would represent very reliable tracers of active outflow
regions.

\begin{figure*}
\includegraphics[height=7cm]{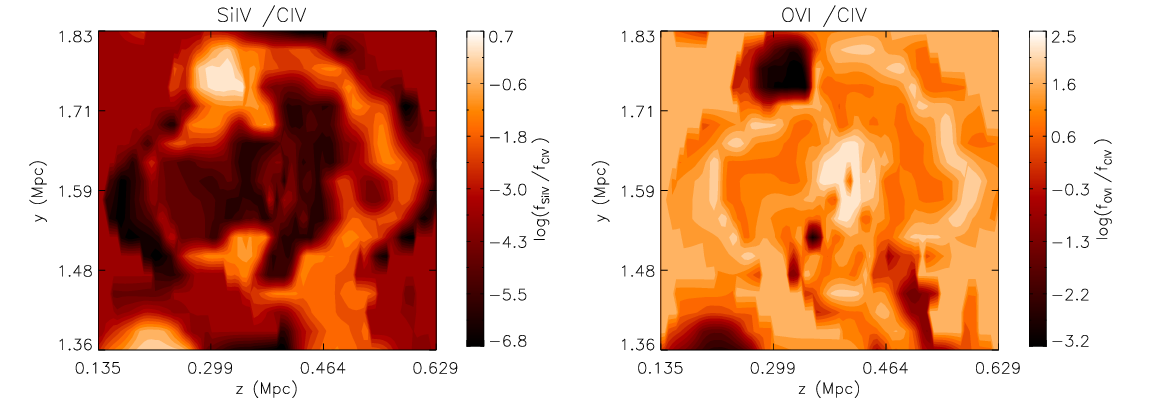}
\caption{Spatial distribution of \SiIV/\CIV$~$(left panel) and
\OVI/\CIV$~$(right) ratios in the simulated volume around the target
galaxy at $z=3.26$. Sizes are in physical units.}
\label{ions_ratios}
\end{figure*}

\begin{figure*}
\includegraphics[height=22cm]{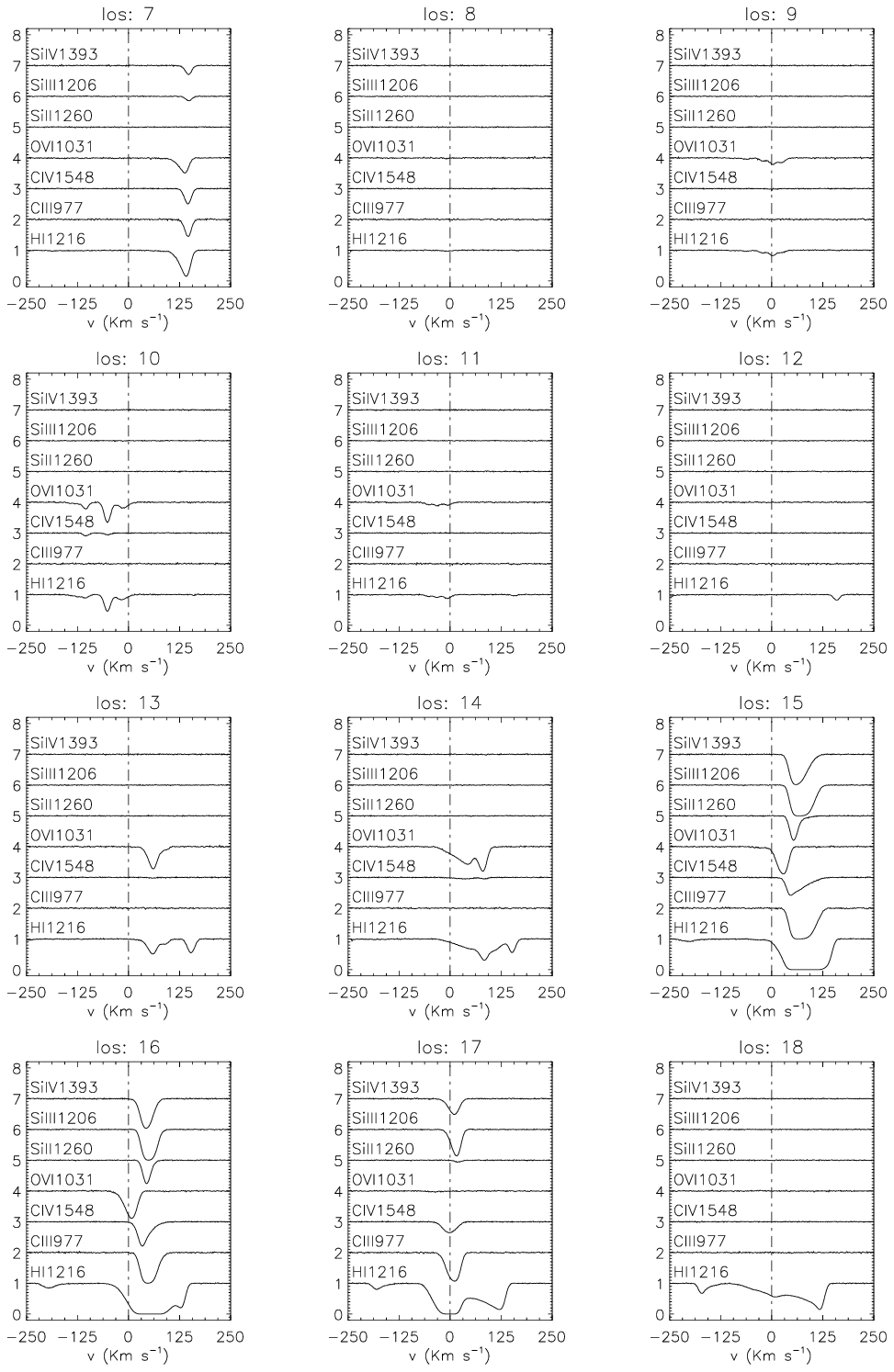}
\caption{Absorption systems for the LOS \#7-18 (labeled at the top
of each panel) through the wind bubble; see Fig. \ref{frac_ion} for
their positions and ionization conditions. Negative (positive)
velocities denote approaching (receding) flows with respect to the
observer.} \label{LOS_bubble}
\end{figure*}

In addition to hot gas (which has a very complex/turbulent velocity
field, resulting in either outflowing or infalling flows) we have
noticed the presence of cooler gas, either in form of infalling
streams or isolated clumps. These cool phases are best traced by
low-ionization species such as \CII, \CIII, \SiII, and \SiIII.  For
example, these ions have enhanced abundances in the cold gas clump
previously identified in the upper left region of the maps.
Intermediate ionization ions (\CIV$~$and \SiIV) are instead
typically found in the interfaces between the cool and the hot gas
phase. It is worth noticing that while \SiIV$~$ has a more patchy
distribution concentrated around the clumps, \CIV$~$extends deeper
into the bubble region and thus into higher temperature regions, as
noticed previously (Fig. \ref{ions_ratios}, left panel). A first
qualitative conclusion drawn from the maps is that the bubbles could
be identified by searching for absorbers that simultaneously have
low \SiIV/\CIV$~$($\log(f_{\rm SiIV}/f_{\rm CIV}) < -2$) and very
high \OVI/\CIV$~$($\log(f_{\rm OVI}/f_{\rm CIV}) > 1$) ratios. These
conclusions are visually supported by the two maps showing the
spatial distribution of these ratios reported in Fig.
\ref{ions_ratios}; typical examples for this behavior are seen in
the spectra obtained from LOS \#10 and \#14 shown in Fig.
\ref{LOS_bubble}, where we present a selection of absorption spectra
through the bubble environment along the LOS defined in Fig.
\ref{frac_ion}. Both LOS exhibit strong \OVI$~$and weak
\CIV$~$absorption systems caused by infalling ($v >0$) material, but
no \SiIV$~$is detected. Despite the higher \CIV$~$ionization
fraction along LOS \#13 (Fig. \ref{frac_ion}) no \CIV$~$absorption
is seen along this sightline (Fig. \ref{LOS_bubble}) due to the very
low gas densities sampled by that  LOS. Stated differently, a
non-detection of \CIV$~$absorption does not necessarily imply a
deficiency of that ion. Instead, its distribution in physical and/or
velocity space could be such that the corresponding absorption is
smeared out. Only sightlines piercing regions of enhanced density
(e.g. LOS \#14) and/or located nearby the target galaxy (LOS \#10)
exhibit clear signs of \CIV$~$absorption.

\begin{figure*}
\includegraphics[width=14cm]{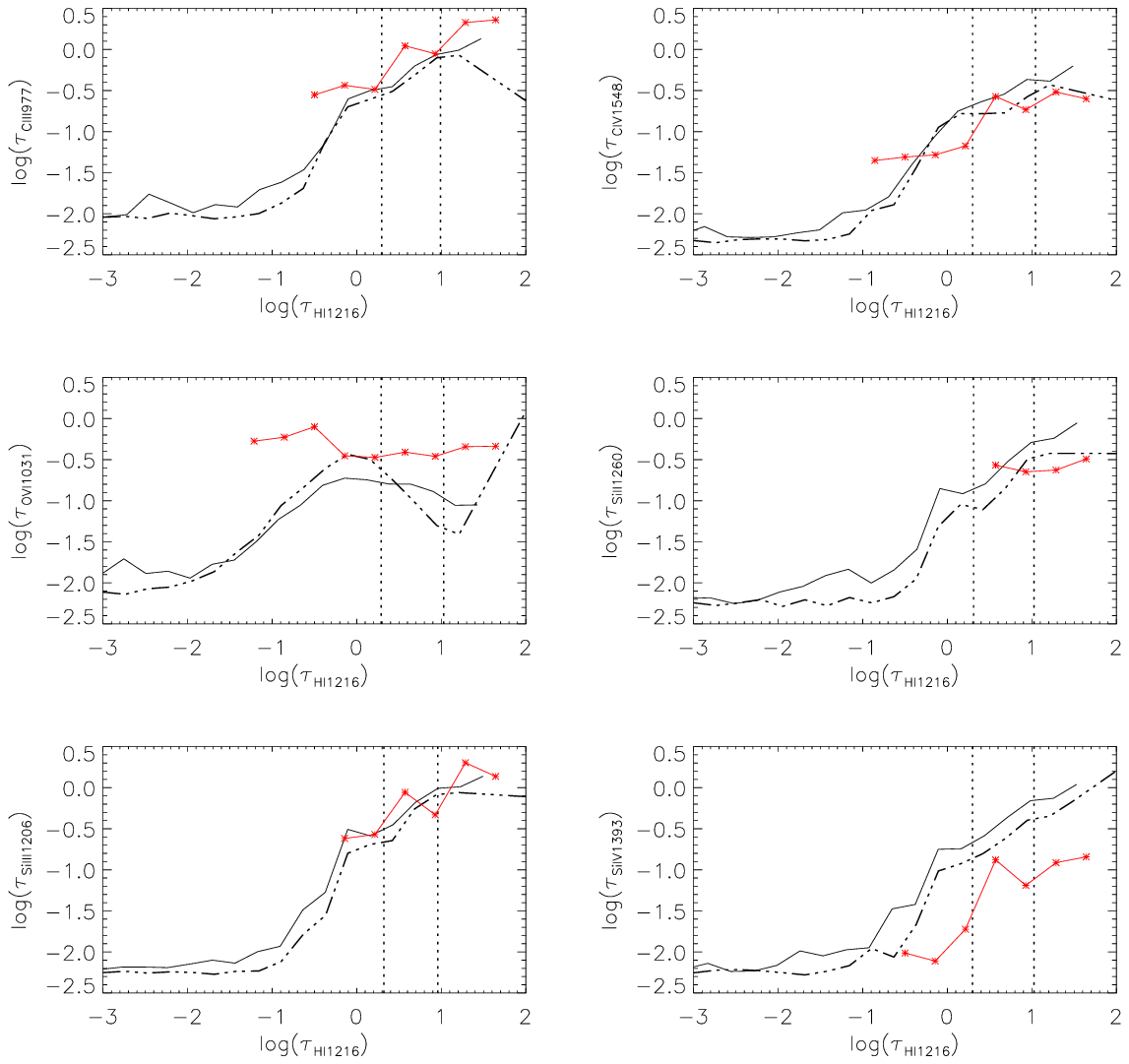}
\caption{Pixel optical depth analysis results for the synthetic
spectra. Plotted are the optical depths of various low- and
high-ionization species against the \HI$~$ \Lya optical depth in the
filament (solid lines) and in the wind (dashed-dot) environment; the
starred line represents observational data from the VLT/UVES LP (see
Tab. \ref{qso-model} and Tab. \ref{qso-test}). The two vertical
lines mark the points at which \Lya (leftmost) or Ly$\beta$
(rightmost) absorption reaches 85$\%$ of the continuum level. Only
bins with more than 50 data points are shown.}
\label{tau_HI_tau_ioni_conv}
\end{figure*}

The cold clump itself instead is characterized by the presence of
\SiII, \SiIII$~$and \SiIV. It appears that the volume occupied by a
given element (here \Si) increases with its ionization state; the
same conclusion holds for carbon ions \CII, \CIII$~$and \CIV.
Typical absorption patterns arising from the clump structures can be
seen in LOS \#15, \#16 and \#17 shown in Fig. \ref{LOS_bubble}.
Absorption features arising from ions with similar ionization states
(e.g., \SiIII$~$and \CIII) are well aligned in radial-velocity
space, although some ions may be more spatially extended than others
(e.g., \CIV$~$vs.\, \SiIV). The contiguity with the hot gas in which
the clump is immersed gives rise to a broad interface region and
causes a \OVI$~$absorption velocity shift of $\Delta v \simeq$ 25 km
s$^{-1}$ with respect to \SiII$~$(LOS \#16). It is worth noticing
that along LOS \#16, piercing the clump, doubly-ionized species
(\CIII$~$and \SiIII) are heavily saturated (see Fig.
\ref{LOS_bubble}).

\subsection{Comparison with the filament environment}

As we have discussed, the target galaxy and its wind bubble are
embedded in a cosmic filament essentially aligned perpendicular to
the plane of the maps shown in Fig. 1, i.e. its symmetry axis is
parallel to the $x$-direction. In order to investigate the spectral
differences between the wind and the filament environments, we have
compared LOS passing through the wind bubble and those intersecting
the filament (hence at an $x$ elevation larger than the bubble
radius $\approx 0.2$ physical Mpc).

To anticipate the main point of the following analysis, we find very
difficult to discriminate cosmic web structures from those created
by the galactic wind purely from available absorption data. The main
reason is that cold clumps resulting from the fragmentation of the
supernova-driven shell have properties that hardly differ from those
of the filament hosting the system.

To further elucidate this point, let us consider a LOS passing through the
bubble on the $y-z$ plane identified by $x=0$; such line will cross
the cool clumps/shell delimiting the hot cavity. Most of the
absorption systems will arise from these dense structures rather
than from the hot, rarefied interior. Thus, although the latter gas
has well defined abundance ratios (e.g. high \OVI/\CIV, low
\SiIV/\CIV$~$as found in the previous analysis of this phase), their
absorption signature is too weak to be detected; on the contrary,
the most evident absorption features come from the cool component.
These will not be different from those obtained from the analysis of
a LOS passing at higher $x$ through the filament only, given that
the physical conditions in the two structures are similar.

It has to be noted that these similar physical conditions originate
from different thermal histories. The cool clumps and shell are the
product of the radiative cooling of the shocked bubble gas from high
temperatures, $T\approx 10^6$~K. In the simulation accretion shocks
are clearly seen in the filaments, too (this fact was first pointed
out by \cite{Cen94}) and these induce complex interface structures
which harbor \OVI$~$and other highly ionized species. Stated
differently, the effect of supernova-induced shock is not easily
disentangled from shocks associated with the gravitational collapse
of small scale structures giving rise to the overdensities we
identify with the \Lya forest\footnote{A similar effect has been
noted by Kawata \& Rauch (2007)}.

A more quantitative conclusion on this issue might come from the POD
analysis of the synthetic spectra presented in Fig.
\ref{tau_HI_tau_ioni_conv}. Shown are the optical depths of various
low- and high-ionization species plotted against the \HI$~$\Lya
optical depth in the filament and in the wind environment. To
correct for saturation, we have complemented the \Lya line with the
weaker Ly$\beta$ and Ly$\gamma$ transitions. The two vertical dotted
lines indicate the saturation point (i.e. when absorption reaches
85\% of the continuum level) for \Lya and Ly$\beta$, respectively,
showing from left to right the regions where these three transitions
were used. To compare our results with observations, we applied the
same analysis to a sample of 28 absorption systems found in 13
spectra from the VLT/UVES Large Program "{\it The cosmic evolution
of the IGM}" (J. Bergeron {\it et al.}) which provides high S/N,
high spectral resolution IGM data for $z\simeq$ 3 (see Tables
\ref{qso-model}, \ref{qso-test}).

Despite the fact that we are tracing high density regions, metals
are detected down to Ly\,$\alpha$ optical depth $\simeq -2.0$. This
is mostly due to the high signal-to-noise ratio [S/N(\CIV)$\simeq$
150, S/N(\OVI)$\simeq$ 98] at the redshift considered.
Unfortunately, a clear distinction in the behavior of the various
ions in the two environments does not emerge: both the filament and
the wind LOS opacities share almost the same trend with the \Lya
optical depth. As expected, \SiII$~$and \OVI$~$are complementary
tracers for the \HI$~$absorption. \SiII$~$traces mostly the
high-density clouds while \OVI$~$is present in lower density
regions. In simulated spectra half of the \HI$~$absorbers, almost
regardless of their physical properties, do not exhibit
\OVI$~$absorption. The existence of these sub-sample lowers the mean
\OVI$~$optical depth and causes the second plateau at high \Lya
optical depth, where the number of high \OVI$~$optical depths points
has a drop. There is a good agreement between our simulated data and
the observations for almost all ions, with the only exception of
\SiIV$~$and, to a lesser extent, of \OVI. This is most probably due
to a selection effect in our sample of absorption systems. For the
\OVI$~$our sample does not represent properly the \OVI-free
absorbers at high \Lya opacities. As for \SiIV, the shift of +0.75
dex seen between predicted opacities and the data is due to the fact
that observed systems are more biased towards intermediate column
density absorbers and they are characterized by very shallow lines;
while in our spectra almost all the \HI$~$absorbers with column
density $\log N_{\rm HI}\simeq 13.8$ are associated with relatively
strong \SiIV$~$absorbers.

\begin{figure*}
\includegraphics[width=14cm]{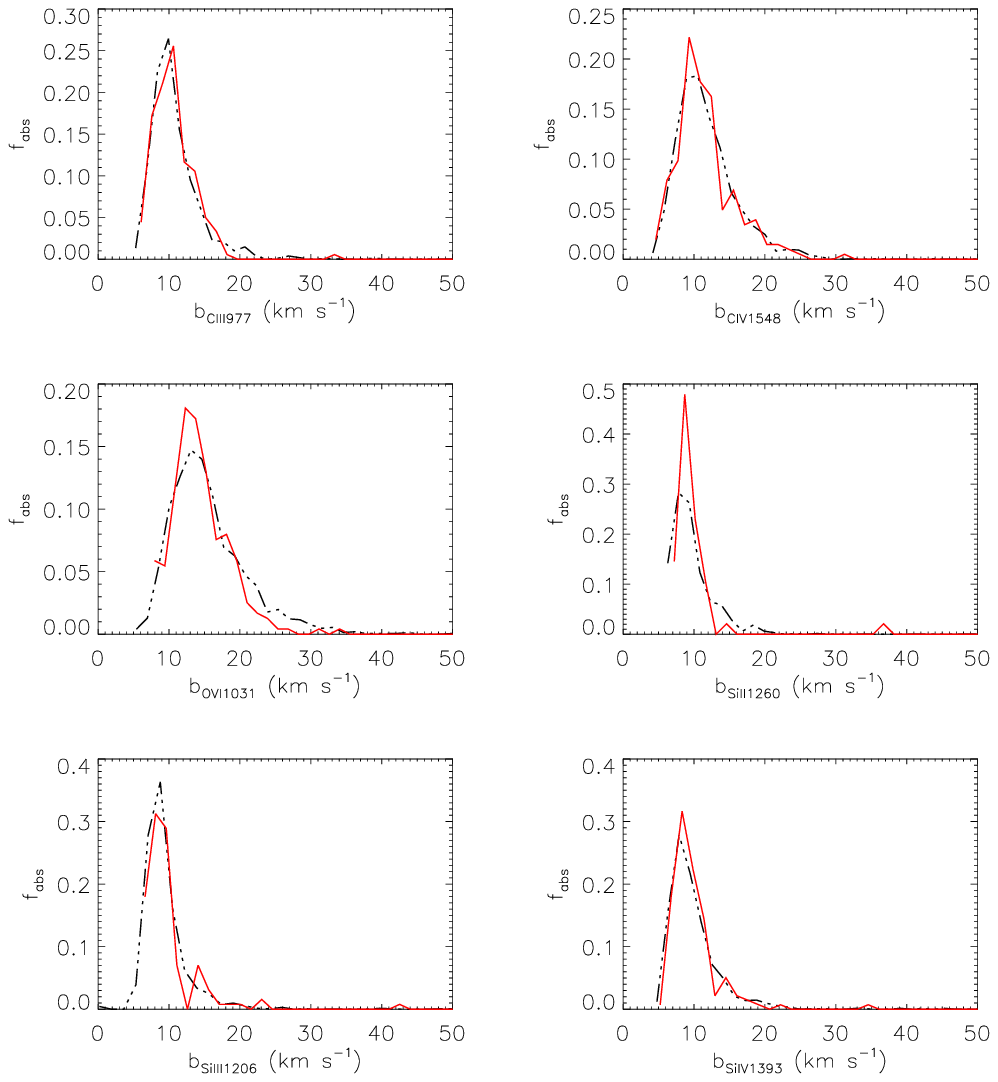}
\caption{Simulated Doppler parameter distribution for the different
ions. Dotted (solid) lines refer to the filament (bubble)
environment. } \label{plot_bH_bM}
\end{figure*}

In addition to the optical depth, the shape and width of the
absorption systems provide additional physical information that
might be in principle used to distinguish between absorbers in the
bubble and in the filament. In fact, these environments are expected
to be dynamically very different. In Fig. \ref{plot_bH_bM} we show
the Doppler parameter distributions for the transitions considered
(as obtained from Voigt profile fitting) in the  filament and bubble
environment. Once again, and for the reasons explained above, no
significant differences between the two distributions emerge for the
considered ions.

\begin{figure*}
\includegraphics[width=14cm]{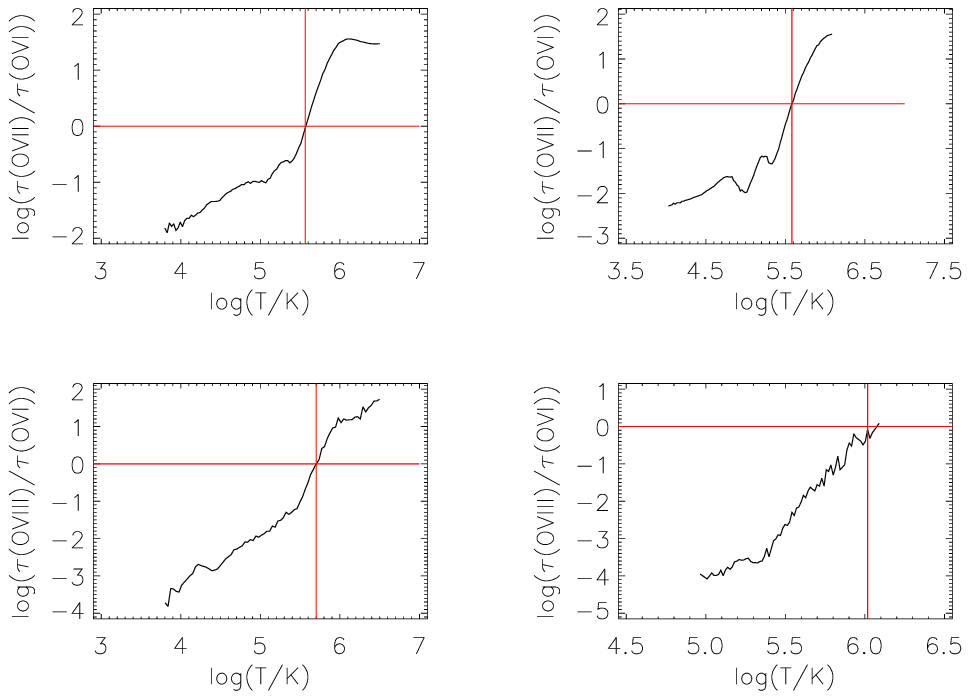}
\caption{Ratios between \OVII$~$and \OVI$~$(upper row), and
\OVIII$~$and \OVI (lower row) optical depths. The left (right)
column refers to the filament (bubble) region. The vertical red
lines indicate the temperature where the two optical depth become
equal.} \label{OVII_OVIII}
\end{figure*}

\subsubsection{Alternative strategies}
As pointed out before, the bubble environment is characterized by
$\log T > 5.5$ and highly ionized species such as \OVI, \OVII$~$and
\OVIII$~$(Fig. \ref{composit_bubble}, \ref{omega_b_T}). Among these
ions, only the \OVI$~$transition falls within the UVES wavelength
range, while those of \OVII$~$($\lambda=21.6$~\AA) and
\OVIII$~$($18.97$~\AA) are located in the X-ray band and will be
observed by future X-ray facilities. It is interesting to briefly
outline the general trends of the optical depth distribution also
for these species.

The optical depth for each transition has been computed using the
following {\it in situ} approximation \citep{Theuns98}: \beq \tau(k)
\simeq n_x(i) \Delta_{px} \frac{(\sigma_0 f\lambda_0)}{1+z} \frac
{c}{\sqrt{\pi} b_x(i)} \Phi_{Gauss}(k), \label{tau_pixel_appr} \eeq
where the symbols are defined after Eq. \ref{tau_abs} and the
physical properties of the gas are taken from the output of the
numerical simulation. To check the sensitivity of the optical depth
to temperature, as a test we have fixed the density to the typical
values found in the bubble and in cool/filament regions, that is
$\log \langle n\rangle \approx -3.7$ and $\log \langle n \rangle
\approx -2.7$, respectively. In Fig. \ref{OVII_OVIII} we show the
predicted \OVII/\OVIII$~$optical depths (normalized to the
corresponding \OVI$~$one) as a function of temperature for both
environments. The optical depths for \OVII$~$and \OVI$~$become equal
for temperatures $\log T\approx 5.6$ in both environments; the ratio
between \OVIII$~$and \OVI$~$ optical depths becomes equal to unity
at higher temperatures in the bubble region (log $T\approx 6.0$)
than in the filament ($\log T\approx 5.7$). Both \OVIII$~$and
\OVII$~$are then expected to provide larger opacities at high
temperatures ($\log T\ge 5.5$) with respect to \OVI, implying that
these species would be particularly well suited to study the hot
component of high redshift galactic winds.

\subsection[]{Additional filament properties}
Finally, we briefly present a statistical analysis of the
\HI$~$Ly$\alpha$ forest in the filament region in which the target
galaxy and its wind bubble are embedded. As the cosmic volume we are
considering is biased by the presence of the relatively massive
target galaxy, we do not expect it to be representative of the
typical low-density environments sampled by Ly$\alpha$ forest
observations. Nevertheless, it is intriguing and instructive to
evaluate the possible differences.

To compare our sample of \HI$~$absorbers with the observed
properties of the Ly$\alpha$ forest in high-redshift spectra we have
used the data published by \citet{Kim_catalog}. These consist of
\Lya absorbers analyzed by Voigt profile fitting of the spectra of 8
QSOs covering the Ly$\alpha$ forest at redshift $1.5 <z< 3.6$
obtained from VLT/UVES observations. For consistency, we have only
used a sub-sample of quasars with $z>3$.
\begin{figure}
\includegraphics[width=9cm]{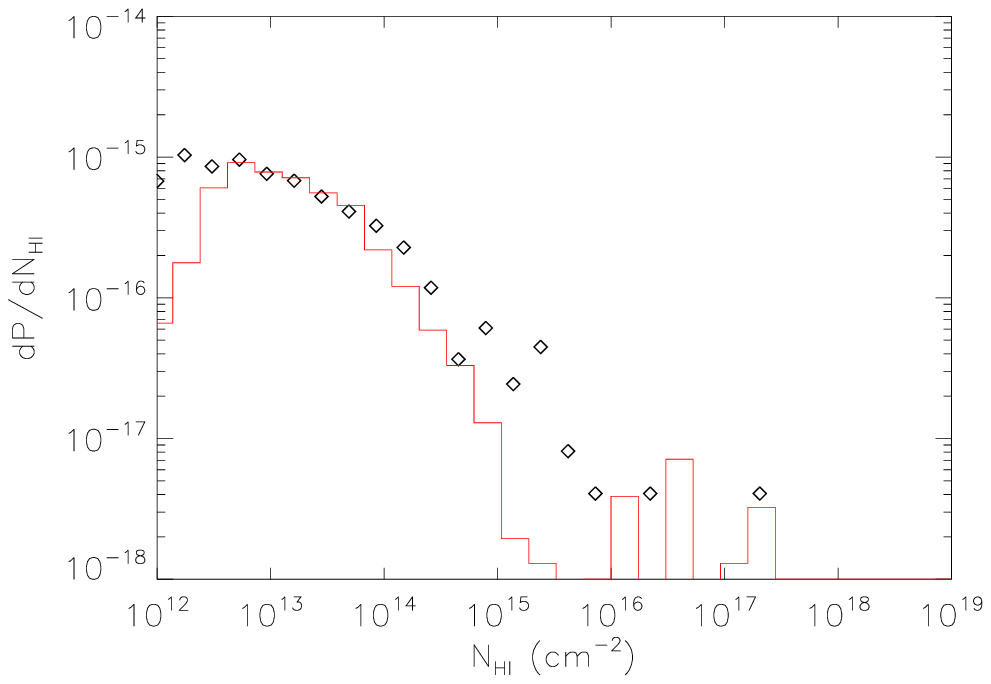}
\caption{Comparison between the \HI$~$column density distribution of
the absorbers in the simulated galaxy environment (red line) and the
observed sample (diamonds). Data are taken from
\citet{Kim_catalog}.} \label{NHI_compar}
\end{figure}
\begin{figure}
\includegraphics[width=9cm]{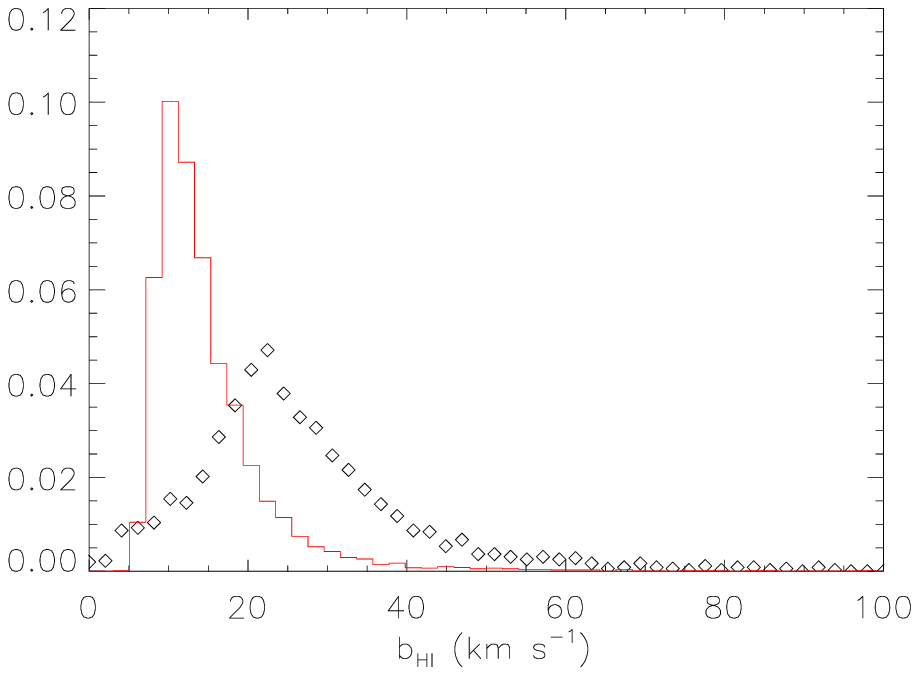}
\caption{Comparison between the \HI$~$Doppler parameter distribution
of the absorbers in the simulated galaxy environment (red line) and
the observed sample (diamonds). The observational data are taken
from \citet{Kim_catalog}.} \label{bHI_compar}
\end{figure}
The results of this comparison in terms of the neutral hydrogen
column density, $N_{\rm HI}$, and Doppler parameter distributions
are shown in Figs. \ref{NHI_compar}-\ref{bHI_compar}. The agreement
in terms of the $N_{\rm HI}$ distribution is generally good in the
range $13 < \log N_{\rm HI} < 15$. The very low column density
deficiency can be possibly caused by the different behavior of the
data analysis software used for the two studies. In
\citet{Kim_catalog} the study was carried using VPFIT (Carswell et
al.) compared to which AUTOVP seems to underestimate the number of
low column density absorbers. Although data incompleteness can also
play a role. As for the Doppler parameter, the mean of the simulated
sample is shifted towards values that are lower than the observed
one. This is plausible if we consider that in our volume we have
large amounts of dense and cold gas (i.e. filament gas) causing
narrow absorption features due to the reduced thermal line
broadening.

\section{Conclusions and discussion}

In this work we have presented an analysis of artificial
absorption-line spectra from the circumgalactic and intergalactic
gas of a wind-blowing galaxy selected from a $z \sim 3$ output from
a SPH simulation. The simulation includes both a multiphase
treatment of the interstellar medium (ISM) and a consistent picture
for the chemical enrichment of the Universe \citep{Brusc03}. With
our feedback scheme we are able to reproduce the observed values for
[\C/\H], [\O/\H] and [\Si/\H] without requiring an overall constant
pollution of the IGM as well as the relation between [\C/\H] versus
overdensity found by \citet{Schaye03}.

The simulated object is comparable in mass with a LBG at this
redshift. The spectra were generated to mimic the behavior of the
UVES spectrograph.

For the wind bubble we find that the interplay between the cooler
infalling material and the outflowing hot gas shapes a bubble-wind
geometry, causing a topology more similar to a pancake like geometry
developing orthogonally to the filament hosting the galaxy, leaving
the filament almost undisturbed. We identify more than four lobes
caused by the cold-bubble material, separated by the so-called
infalling-channels. If this is typical for wind-blowing galaxies at
high redshift, then the gaseous environment is determined by: (i)
the cold filament hosting the galaxy, (ii) a mixture of hot gas ($T
> 10^{5}$ K) filling up the cavity created by the galactic wind,
(iii) the almost pristine IGM infalling towards the galaxy and
opposing to the hot-bubble expansion, (iv) the colder blown out
material that could have had the time to fragment into smaller
clumps.

>From the spectral analysis we find that cold-bubble clumps are
characterized by \SiII,$~$\SiIII,$~$\CII$~$and \CIII$~$absorption.
High oxygen ions lines are not expected in these clumps, but the
cooler gas (as indicated by \SiII$~$absorption) is enveloped by an
\OVI$~$phase.

The outflowing gas from the galaxy is characterized by the high
ionization states of oxygen, i.e., \OVI, \OVII, and \OVIII. While
only \OVI$~$is observable in UV/optical absorption, we show that the
\OVII$~$and the \OVIII$~$phase should give rise to absorption with
substantial optical depths in the X-ray band.

A physical environment similar to that of the cold clumps within the
bubble is present in cooler regions in the surrounding intergalactic
filament (i). Here, the cold and warm filamentary material produces
low- and high ion absorption patterns that mimic those found in the
bubble environment. From an analysis of the optical depth and
velocity width distribution of the absorption in both environments
we find that there is no clear way to distinguish between these two
spatially very distinct environments based on their absorption-line
patterns.

The comparison of our optical depths with the observed data shows a
good agreement for almost all ions except for \OVI$~$and \SiIV. This
is most likely due to the composition of our observed sample. In our
simulated spectra, half of the LOSs showing strong \Lya optical
depths do not show appreciable \OVI$~$absorption; while our observed
sample lacks in these systems causing a higher mean \OVI$~$optical
depth. As concerning \SiIV, our observed sample is mostly composed
by intermediate column density absorbers. In our simulated spectra
instead, higher \HI$~$column densities absorbers are preferably
associated with medium/high column densities \SiIV$~$absorbers.

Our study suggests that it is quite difficult to discriminate
between the various circumgalactic and intergalactic gaseous phases
of a starbursting galaxy environment based solely on the absorption
line characteristics. Enlarging the sample of target objects would
probably increase the statistical weight of our conclusions, but we
do expect that the general picture is preserved. The incidence of
cold clumps on the spectra is expected to increase in more complex
environment, as for example if more filaments are intersecting each
other. Along with a different metal distribution, bubbles will be
instead characterized by almost the same range of temperatures and
densities; they will then give raise to very similar absorption
patterns. Metal cooling can then be the most important missing
ingredient of the present study. Non-equilibrium ionization effects
could also leave an imprint on the cooled gas, which might still
keep memory of the past hot and rarefied condition. We are planning
to explore these variables on future works.

$~$

Finally, a brief comparison with two recently appeared studies
(\cite{Oppenheimer06}; \cite{Kawata07}) must be given.

In \citet{Oppenheimer06} the authors mainly use statistics related
to \CIV$~$ absorption to compare their different feedback recipes
with observational result. Consistently with their findings, we
confirm that most of the \CIV$~$ is of collisional origin; however,
our results also reproduce the [\C/\H] vs. overdensity relation by
\citet{Schaye03}. Most probably, this discrepancy is due to a
different metallicity distribution. \citet{Oppenheimer06} predict a
metallicity $\simeq$ 1 dex higher than our values; also their
fiducial models show an almost flat pollution at high overdensities
(see their Fig. 10) contrary to our continuously rising level.
Still, both works predict that winds expand preferably in low
density regions, leaving the filaments hosting the galaxy
unaffected.

As for \citet{Kawata07}, their main finding is that \OVI$~$is the
ion most sensible feedback tracer. Their assumption of a constant -
and somewhat unphysical - IGM pollution does not allow a direct
comparison with our metallicity distribution. Nevertheless, by
looking at their maps it seems that the bulk of the \OVI~absorption
arises from shocked filament material as well as from clumps
condensed from blown out gas, a scenario in agreement with the one
proposed here. Repeating their search for strong \OVI$~$absorbers
(with flux $>$ 0.8) associated with unsaturated \HI$~$ones, we find
that these systems are almost evenly produced in the bubble
environment and the filament one. We cannot then confirm the
exclusive association of these systems to winds.

\appendix

\section[]{Synthetic Spectra}

\subsection{Smoothing}

\begin{figure}
\includegraphics[width=8.5cm]{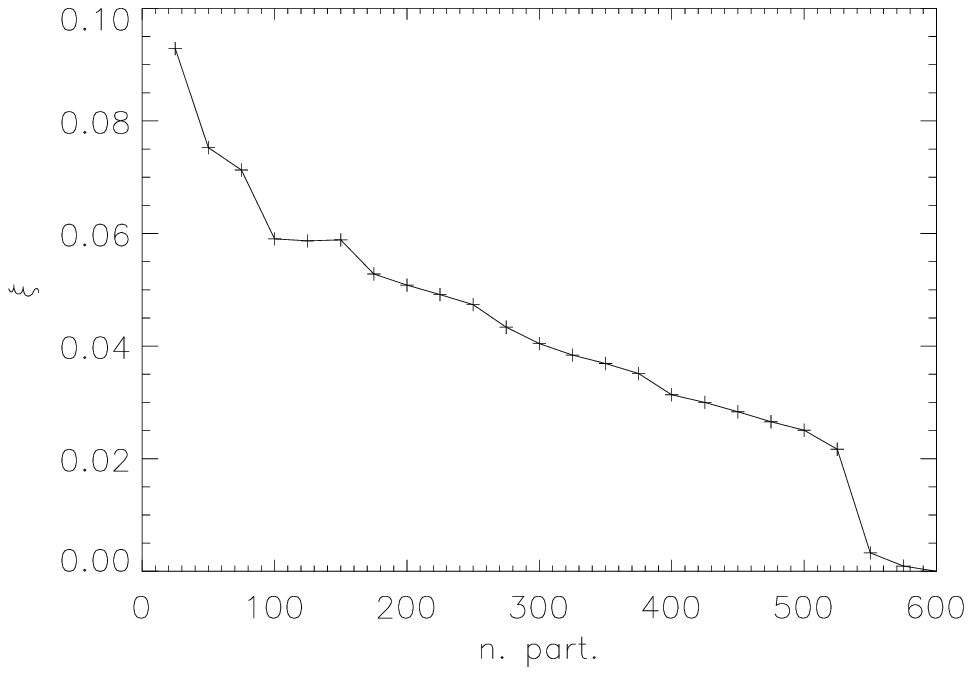}
\caption{Smoothing convergency test using a gaussian kernel with a FWHM $\sigma =$ 8.0 {\rm h$^{-1}$} kpc.}
\label{convergency}
\end{figure}

To calculate a spectrum along a given line of sight (LOS) through
the simulation box it is necessary to recover the physical
properties of the gas at each point along the path, by smoothing the
various fields. For this, the line of sight has been divided into
cells, to which the values of the gas density, temperature,
velocity, and metallicity are assigned by a weighted mean on the
surrounding particles. The weight of each particle is determined by
its distance from the LOS-cell using a specified kernel function.
For consistency with the adopted SPH simulation, we have chosen a
gaussian kernel with FWHM $\sigma =$ 8.0 {\rm h$^{-1}$} kpc
comoving. The number of particles used to derive the mean is fixed
by the balance between convergence and computational expense. To
determine such number, we have recovered the various physical
quantities for 600 cells along the LOS using a variable number of
particles (up to 600, the reference case). In Fig. \ref{convergency}
we plot the results of such test in terms of the quantity \beq \xi_j
=\langle\frac{|X_i^j-X_{600}^j|}{X_{600}^j}\rangle_j\,
\label{zeta_conv} \eeq where $X_i^j$ indicates the $j$-th physical
quantity recovered using $i$ particles, and the mean is calculated
over the cells considered. As we see, an accuracy of $\la 6\%$ is
obtained by using 100 particles; by further taking into account the
computational expense, we have fixed this value as the optimal one
and have adopted it in the following.

\begin{figure*}
\includegraphics[height=5.5cm,width=\textwidth]{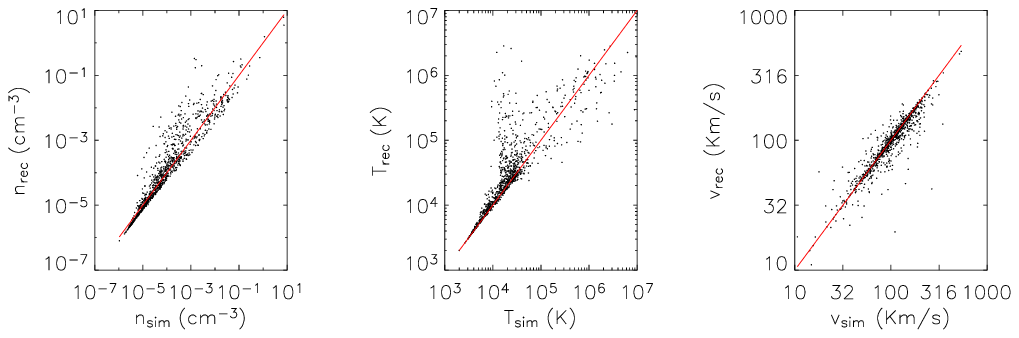}
\caption{Smoothing recovery test for the density (left panel), temperature (middle) and velocity (right) fields.
The dots represent the relation between the recovered and the simulated physical properties of 1000 particles in
the high resolution simulation at redshift $z \sim 3.0$. The line represents the case of an ideal recovery algorithm.}
\label{recover_scatter}
\end{figure*}
To further test the smoothing algorithm, we have recovered the physical properties of 1000 gas particles.
Fig. \ref{recover_scatter} shows a comparison between recovered and simulated values. The accuracy of recovered quantities is
reasonably good. The scatter in the distribution may be related to particles belonging to regions
near the interfaces of virializing halos, the sharp transition across the shock front
characterizing these regions been smoothed away by the recovery process. This interpretation is
supported by the ``wall" in the temperature distribution at $T \simeq 10^4 K$, i.e. the lowest
temperature allowed by the adopted atomic cooling function. Moreover, the small filling factor of
virialized regions tends to translate into a density overestimate. The velocity distribution
seems instead more symmetric, as a result of the different kinetic behavior characterizing the
collapsed haloes distribution.
The metallicity field is particularly affected by the recovery process, as expected from the fact that no diffusion
equations are solved for the heavy elements. As a result, polluted and pristine regions are
separated by unphysical sharp boundaries. Results concerning the metallicity derived from our
synthetic spectra should be taken with caution.

\subsection{UVES models}
For our synthetic spectra we have chosen to mimic the behavior of the UVES spectrograph
mounted at the ESO Very Large Telescope (VLT). To obtain a reliable model for the signal to noise ($S/N$) and for the signal
to error ($S/E$) ratio we have used a set of observations made as part of the Large Program
"{\it The cosmic evolution of the IGM}" (J. Bergeron {\it et al.}) which provides high S/N, high
spectral resolution IGM data for $z\simeq$ 3 (Tab. \ref{qso-model}).
\begin{table}
\begin{center}
\begin{tabular}{lcl}
\hline
\multicolumn{1}{c}{z}&\multicolumn{1}{c}{QSO}\\
\hline
\hline
$2.170$&${\rm Q0122-380}$\\
$2.208$&${\rm PKS1448-232}$\\
$2.280$&${\rm HE0001-2340}$\\
$2.406$&${\rm Q0109-3518}$\\
$2.406$&${\rm HE2217-2818}$\\
$2.434$&${\rm HE1347-2457}$\\
$2.740$&${\rm HE0151-4326}$\\
$2.758$&${\rm Q0002-422}$\\
\hline
\end{tabular}
\caption{QSOs spectra used to compute the models for $S/N$ and $S/E$ ratios for the UVES
spectrograph.} \label{qso-model}
\end{center}
\end{table}

\begin{table}
\begin{center}
\begin{tabular}{lcll}
\hline
\multicolumn{1}{c}{z}&\multicolumn{1}{c}{QSO}\\
\hline
\hline
$2.134$& ${\rm HE1341-1020}$\\
$2.685$& ${\rm PKS0329-255}$\\
$2.661$& ${\rm Q0453-423}$\\
$2.885$& ${\rm HE2347-4342}$\\
$3.054$& ${\rm HE0940-1050}$\\
\hline
\end{tabular}
\caption{QSOs spectra used to test the models for $S/N$ and $S/E$ ratios for the UVES
spectrograph.} \label{qso-test}
\end{center}
\end{table}
Considering that the maximum restframe absorption wavelength used in
our spectra is $\lambda \simeq 1550$ \AA, we have fixed the
wavelength range for our models to $ 3300< \lambda <7400$ \AA.

\subsubsection{$S/N$ model}
\begin{figure}
\includegraphics[width=9cm]{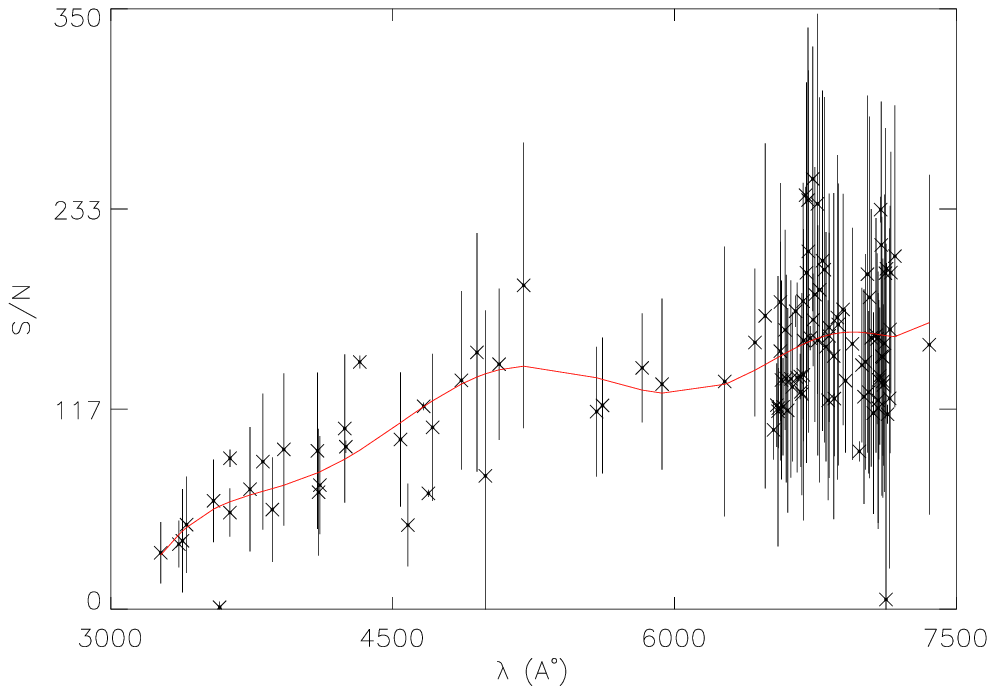}
\caption{Fiducial model of the $S/N$ for the UVES spectrograph. In black we show the observed data, along with their variance; while the red line refers to our polynomial fit. We can note the different density of data with $\lambda$, this reflects the varying number density of the absorbing clouds.}
\label{S_N_model}
\end{figure}

\begin{figure}
\includegraphics[width=9cm]{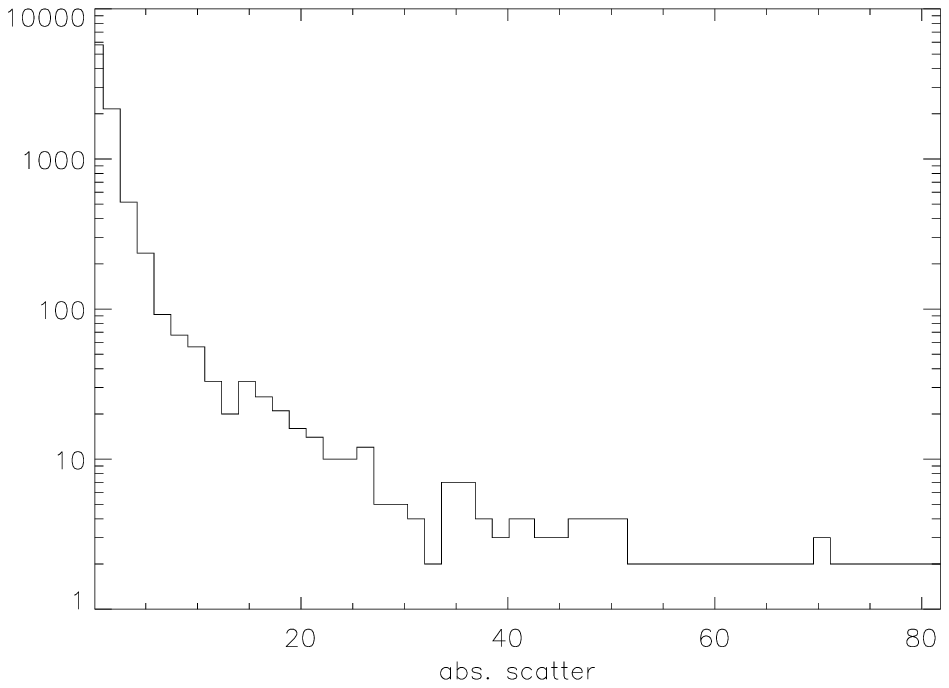}
\caption{Test for the $S/N$ model (Fig. \ref{S_N_model}). The steep decrease in the scatter distribution indicates a satisfactory model behavior.}
\label{S_N_test}
\end{figure}

To obtain the $S/N$ dependence with wavelength we have searched (by
eye) for unabsorbed regions in the UVES spectra listed in Tab.
\ref{qso-model}. The data obtained then were binned in 0.25 \AA$~$
bins, so that every bin is formed by 10 pixels at the UVES
resolution of $\simeq 0.025$ \AA$~$ per pixel. We then have
considered the mean value as estimate for the signal from each
pixel/bin, the noise was estimated by a simple subtraction. The
triplet of values ($\lambda$-$S$-$N$) obtained in this way were
collected in a single large sample sorted by wavelengths. To reduce
the data used for the fitting, we have divided the final data set in
bins of 100 values. For each bin we then have estimated the mean
value and the scatter for the $S/N$ by fitting the distribution of
values using a gaussian. The values obtained were then fitted using
a polynomial function (Fig. \ref{S_N_model}).
In Fig. \ref{S_N_test} we show the test of our fiducial model for the $S/N$ using the QSOs given in Tab. \ref{qso-test}. We have plotted the distribution of the absolute scatter between predicted and observed $S/N$\footnote{The data were treated in the same way as for the model.}. As we can see, also if the distribution reaches high values; the rapid decrease of the observed histogram satisfies our model.
\subsubsection{$S/E$ model}
\begin{figure}
\includegraphics[width=9cm]{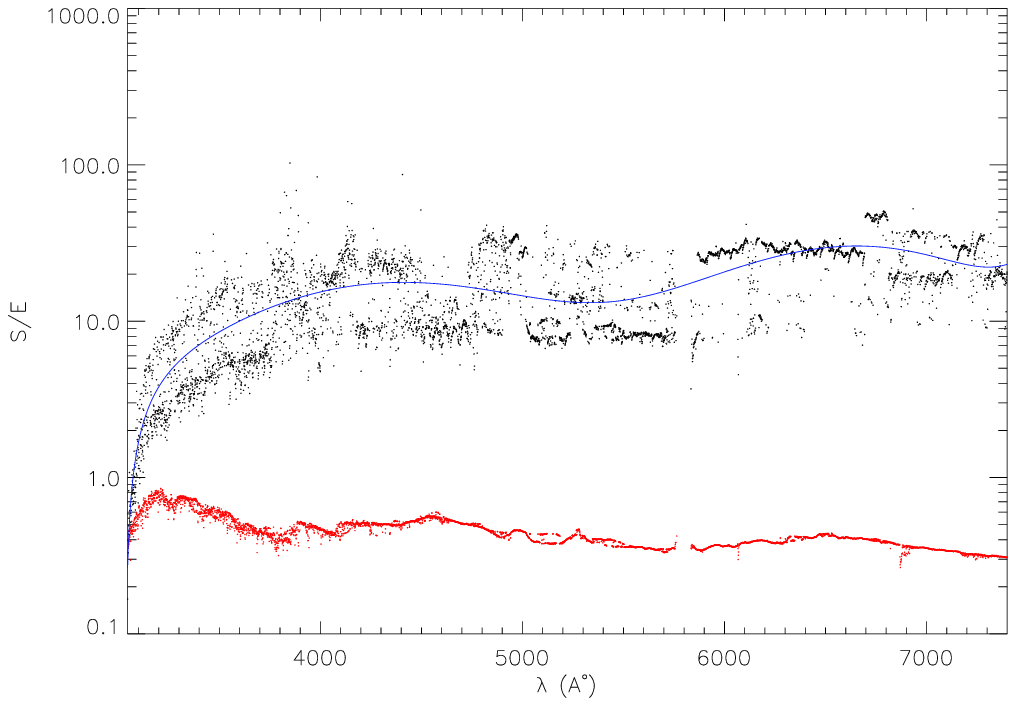}
\caption{Dependence of the $S/E$ with wavelength. The black dots refer to the observed data, while the blue line represents our polynomial fit. Red dots represent the predicted ratios using the Poissonian approximation.}
\label{S_E_model}
\end{figure}

To obtain a model for the signal-to-error ratio, we have treated the observed data as described in the previous section. In Fig. \ref{S_E_model} we show the observed data (black dots) along with our model (blue dots) and a Poissonian model (red dots).
We see how the latter approximation (usually considered) overestimates the errors associated with fluxes. \begin{figure}
\includegraphics[width=9cm]{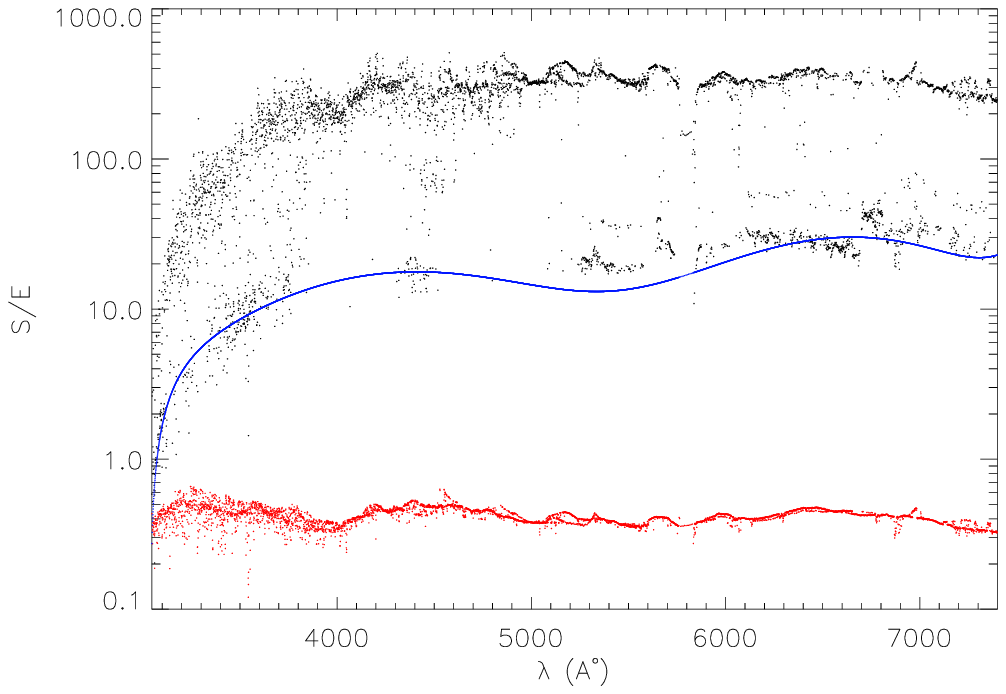}
\caption{Test for our $S/E$ model. The colors are as in Fig. \ref{S_E_model}.}
\label{S_E_test}
\end{figure}
>From Fig. \ref{S_E_test} we can see that both the approximations
overestimate the errors associated with fluxes. However, our model
can be considered as a better low-limit approximation to the real
ratio.

\subsubsection{Binning issues}

\begin{figure*}
\includegraphics[height=5.5cm]{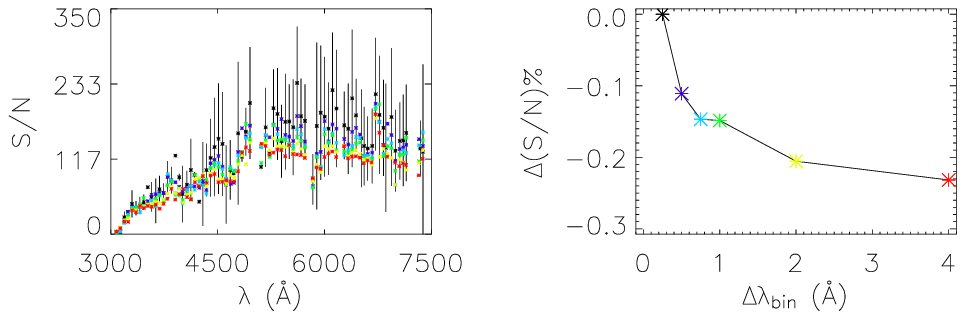}
\caption{Dependencies of the $S/N$ model on the binning size. On the
left we show the observed $S/N$ ratio for different wavelengths.
Colors, from black to red, refer to a growing binning size (0.25,
0.50, 0.75, 1.0, 2.0, 4.0 \AA). For the first binning size (our
working size) we show also the relative errors. The right plot shows
the observed mean relative scatter of each set from the reference
one as a function of the binning size.} \label{S_N_rebin}
\end{figure*}

To assess the dependencies of our $S/N$ model, we have carried out
the binning procedure using different bin sizes. In Fig.
\ref{S_N_rebin} we show on the left our results for binning sizes of
0.25, 0.50, 0.75, 1.0, 2.0, 4.0 \AA , respectively, as dots with
colors from black to red. The right plot shows the mean relative
scatter of each data set, from the reference one (that relative to a
0.25 \AA$~$binning).

These two plots demonstrate that the main consequence of rebinning
is a lowering of the observed $S/N$ ratio, most probably due to the
intrinsic continuum variation inside the bin. Since we do not have
fully analyzed spectra, and considering that every new set of $S/N$
data lies into the error bars of our reference-data set, we can
consider our model as acceptable.

\subsection{Spectrum generation and degradation}

To obtain the absorption arising along the LOS, that means the
optical depth $\tau$ for each cell, we have used the Eq.
\ref{tau_abs} \citep{Theuns98}: \beq \label{tau_abs}
\tau(k)=\sum_{i=1}^{N_{px}} n_x(i) \Delta_{px} \frac{(\sigma_0 f
\lambda_0)}{1+z} \frac {c}{\sqrt{\pi} b_x(i)} \Phi_{Voigt}(k, i)
\eeq \beq n_x(i)=(Z_{II}(i) Y_{II,x}(i) + Z_{III}(i) Y_{III,x}(i))
n(i) f(A_x, \mathbf{\chi}_+) \eeq where:
\begin{enumerate}
\item $k$ is the cell where we want to calculate the optical depth;
\item the summation is over all the cells, $N_{px}$, in each LOS;
\item $n_x(i)$ is the numeric density in the cell $i$ of the element $x$ with ionization degree $\chi_+$;
\item $n(i)$ is the total gas density in the cell $i$;
\item $Y_{II,x}$ and $Y_{III,x}$ are the yields for the element $x$, respectively for PopII and PopIII stars (Tab. \ref{yields});
\item $f(A_x, \chi_+)$ is the ionization fraction;
\item $\Delta_{px}$ is the linear dimension of the cell;
\item $z$ is the redshift;
\item $\sigma_0,\;f,\;\lambda_0$ are atomic constants, respectively: $\sigma_0$ is linked to the Thomson cross section, $\sigma_T$, as follow $\sigma_0 = \sqrt{\frac{3 \pi}{8} \sigma_T}$, and $f$ is the oscillator strength for the transition $\lambda_0$\footnote{The atomic parameters for each transition were taken from \citet{verner}.};
\item $b_x$ is the Doppler parameter for the element $x$ defined as:
\beq \label{v_termica} b_x(i)=\sqrt{\frac{2kT(i)}{m_x}} \simeq
12.845 \sqrt{\frac{T_4(i)}{A_x}}\;{\rm km\;s^{-1}} \eeq where: $k$
is the Boltzmann constant, $T(i)$ is the temperature of the i-th
cell in Kelvin\footnote{$T_4 \equiv T/10^4$ K.}, $m_x$ and $A_x$ are
respectively the mass and the atomic number of the element $x$;
\item $\Phi_{Voigt}(k, i)$ is the statistical weight that determine the contribution of the cell i-th to the absorption of the cell $k$-th, in our case we have chosen to use a Voigt profile.
\end{enumerate}

\begin{table}
\begin{center}
\begin{tabular}{lllll}
\hline \multicolumn{1}{c}{Element}& \multicolumn{1}
{c}{PopIII}&\multicolumn{1}{c}{PopII}\\
\hline \hline
$\C$& $0.0488$& $0.109$\\
$\O$& $0.504$& $0.693$\\
$\Si$& $0.211$& $0.0626$\\
\hline
\end{tabular}
\caption{Yields for the element included in our spectra (\citet{Woosley95}, \citet{Heger02}).}
\label{yields}
\end{center}
\end{table}
The absorption wavelength associated to each $\tau(k)$ is obtained
considering the rest-frame absorption wavelength at the redshift of
the numerical output\footnote{The redshift difference between two
cells is less than 0.01\%.} and the total velocity of the cell. To
compare with the data, it is necessary to degrade the raw spectrum
to account for finite instrumental resolution and noise. To this aim
we convolve the spectra with the (UVES) instrumental profile assumed
to be a gaussian of width $\simeq$ 6.6 km s$^{-1}$. The noise
contribution to the flux in each bin is obtained by considering a
gaussian noise with zero mean and with width determined by our S/N
model. The same is done for the error associated with each flux
value. In Tables \ref{metal_lines} and \ref{HI_lines} we list the
atomic parameters for the atomic transition included in our
simulated spectra. Here we show respectively in the first column the designation for the
absorption followed by the rest frame absorption wavelength,
oscillator strength, gamma value and the second absorption
wavelength in case of a doublet.
\begin{table*}
\begin{center}
\begin{tabular}{lllllllllllll}
\hline
\multicolumn{1}{c}{Ion}& \vline &\multicolumn{1}{c}{$\lambda_1 ($\AA$)$}& \vline &\multicolumn{1}{c}{f}& \vline &\multicolumn{1}{c}{$\gamma/$10$^9$}& \vline &\multicolumn{1}{c}{$\lambda_2 ($\AA$)$}\\
\hline
\hline
$\CIV 1548$& $ \vline $& $1548.195$& $ \vline $& $0.190$& $ \vline $& $0.2650$& $ \vline $& $1550.770$\\
$\OI 1302$& $ \vline $& $1302.168$& $ \vline $& $0.0504$& $ \vline $& $0.5750$& $ \vline $& $-$\\
$\OVI 1031$& $ \vline $& $1031.926$& $ \vline $& $0.133$& $ \vline $& $0.4.125$& $ \vline $& $1037.617$\\
$\SiII 1526$& $ \vline $& $1526.707$& $ \vline $& $0.132$& $ \vline $& $1.9600$& $ \vline $& $-$\\
$\SiII 1304$& $ \vline $& $1304.370$& $ \vline $& $0.0871$& $ \vline $& $1.7200$& $ \vline $& $-$\\
$\SiII 1260$& $ \vline $& $1260.422$& $ \vline $& $1.180$& $ \vline $& $2.5330$& $ \vline $& $-$\\
$\SiII 1193$& $ \vline $& $1193.290$& $ \vline $& $0.584$& $ \vline $& $3.4950$& $ \vline $& $-$\\
$\SiII 1190$& $ \vline $& $1190.416$& $ \vline $& $0.293$& $ \vline $& $3.5030$& $ \vline $& $-$\\
$\SiIII 1206$& $ \vline $& $1206.500$& $ \vline $& $0.0168$& $ \vline $& $2.550$& $ \vline $& $-$\\
$\SiIV 1393$& $ \vline $& $1393.755$& $ \vline $& $0.524$& $ \vline $& $0.88250$& $ \vline $& $1402.770$\\
\hline
\end{tabular}
\caption{Atomic parameters for the metals transitions included in our simulated spectra \citep{verner}. $\lambda_2$ refers to the second component in doublets.}
\label{metal_lines}
\end{center}
\end{table*}

\begin{table*}
\begin{center}
\begin{tabular}{llllllllllll}
\hline
\multicolumn{1}{c}{Ion}& \vline &\multicolumn{1}{c}{$\lambda_1 ($\AA$)$}& \vline &\multicolumn{1}{c}{f}& \vline &\multicolumn{1}{c}{$\gamma/10^7$}\\
\hline
\hline
$\HI 1216$& $ \vline $& $1215.670$& $ \vline $& $0.416$& $ \vline $& $62.650$\\
$\HI 1026$& $ \vline $& $1025.722$& $ \vline $& $0.0791$& $ \vline $& $18.970$\\
$\HI 972$& $ \vline $& $972.537$& $ \vline $& $ 0.0290$& $ \vline $& $8.126$\\
$\HI 950$& $ \vline $& $949.743$& $ \vline $& $0.0139$& $ \vline $& $7.640$\\
$\HI 938$& $ \vline $& $937.804$& $ \vline $& $0.780\cdot10^{-2}$& $ \vline $& $4.4230$\\
$\HI 931$& $ \vline $& $930.748$& $ \vline $& $0.481\cdot10^{-2}$& $ \vline $& $1.236$\\
$\HI 926$& $ \vline $& $926.226$& $ \vline $& $0.318\cdot10^{-2}$& $ \vline $& $0.8249$\\
$\HI 923$& $ \vline $& $923.150$& $ \vline $& $0.222\cdot10^{-2}$& $ \vline $& $0.5782$\\
$\HI 921$& $ \vline $& $920.963$& $ \vline $& $0.160\cdot10^{-2}$& $ \vline $& $0.4208$\\
$\HI 919$& $ \vline $& $919.351$& $ \vline $& $0.120\cdot10^{-2}$& $ \vline $& $0.3158$\\
$\HI 918$& $ \vline $& $918.129$& $ \vline $& $0.921\cdot10^{-3}$& $ \vline $& $0.2430$\\
$\HI 917$& $ \vline $& $917.181$& $ \vline $& $0.723\cdot10^{-3}$& $ \vline $& $0.1910$\\
$\HI 916$& $ \vline $& $916.429$& $ \vline $& $0.577\cdot10^{-3}$& $ \vline $& $0.1529$\\
\hline
\end{tabular}
\caption{Atomic parameters for the \HI$~$included in our simulated spectra \citep{verner}.}
\label{HI_lines}
\end{center}
\end{table*}

\section[]{Sample spectra}
In the following Figs. \ref{spectra_bubble_HI} - \ref{spectra_bubble_SiIV} we show the
simulated spectra along a slice of the simulated volume containing the target galaxy.
\clearpage
\begin{figure*}
\includegraphics[height=22cm]{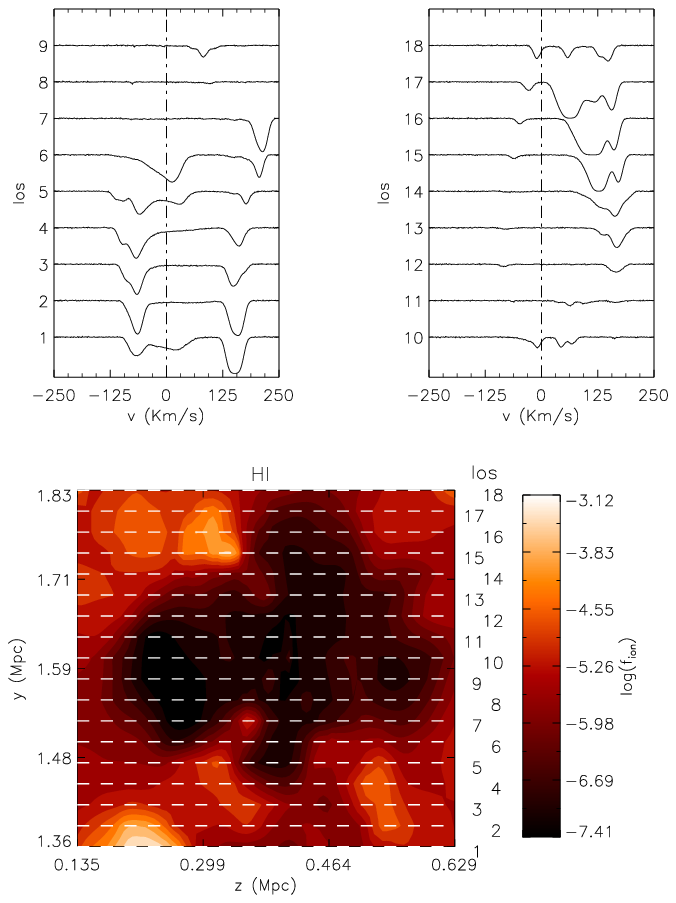}
\caption{The map shows for the ionization fraction of the ion
indicated by the label. The dotted lines identify by progressive
number the LOS whose spectra are shown in the upper panels, For ions
not discussed in detail, we show in the upper panel the expected
column density derived from an integration along each LOS.}
\label{spectra_bubble_HI}
\end{figure*}
\clearpage
\begin{figure*}
\includegraphics[height=22cm]{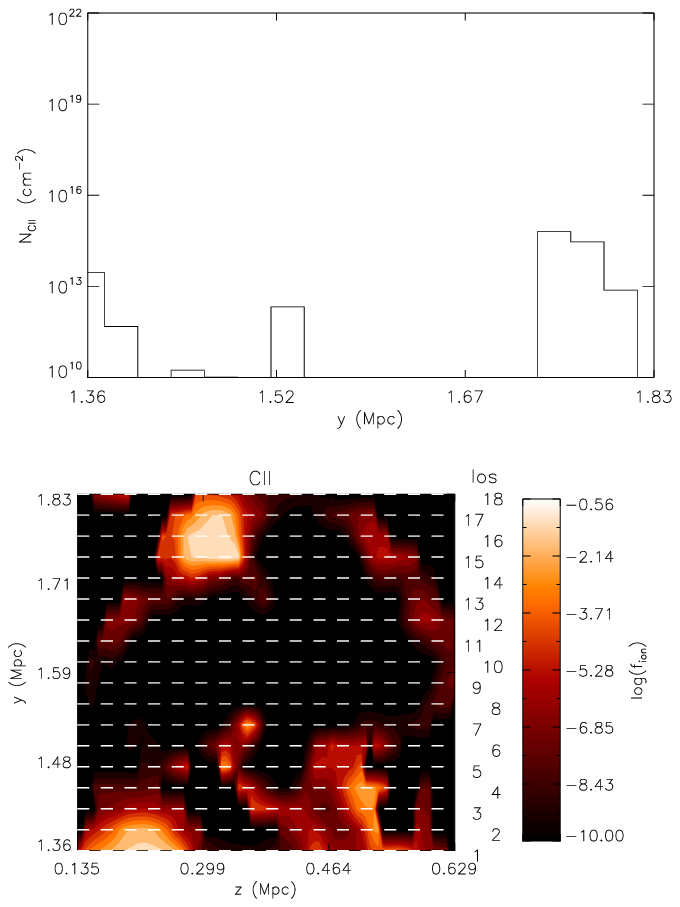}
\caption{} \label{spectra_bubble_CII}
\end{figure*}
\clearpage
\begin{figure*}
\includegraphics[height=22cm]{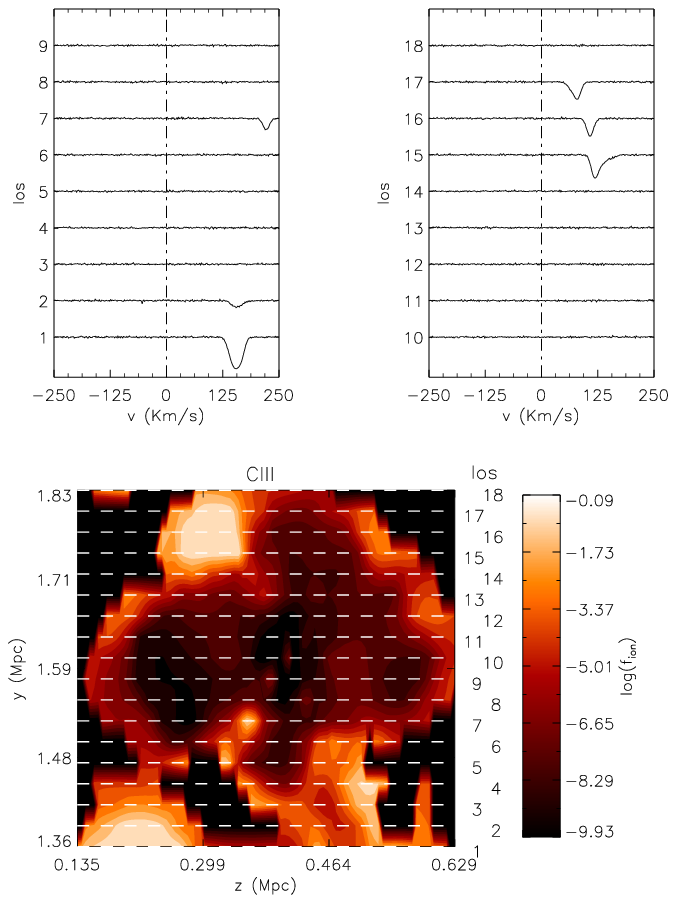}
\caption{} \label{spectra_bubble_CIII}
\end{figure*}
\clearpage
\begin{figure*}
\includegraphics[height=22cm]{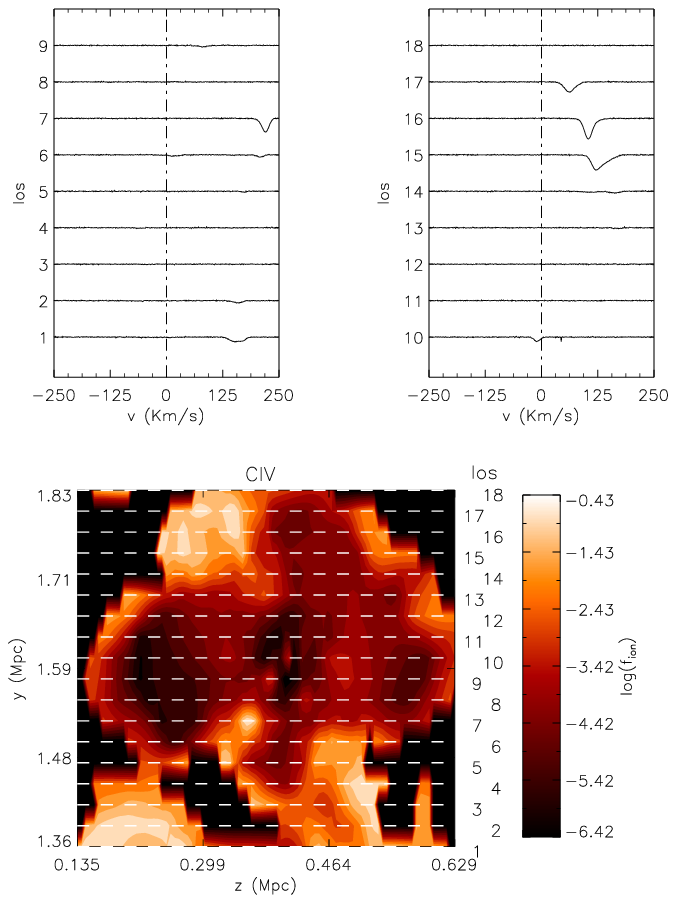}
\caption{} \label{spectra_bubble_CIV}
\end{figure*}
\clearpage
\begin{figure*}
\includegraphics[height=22cm]{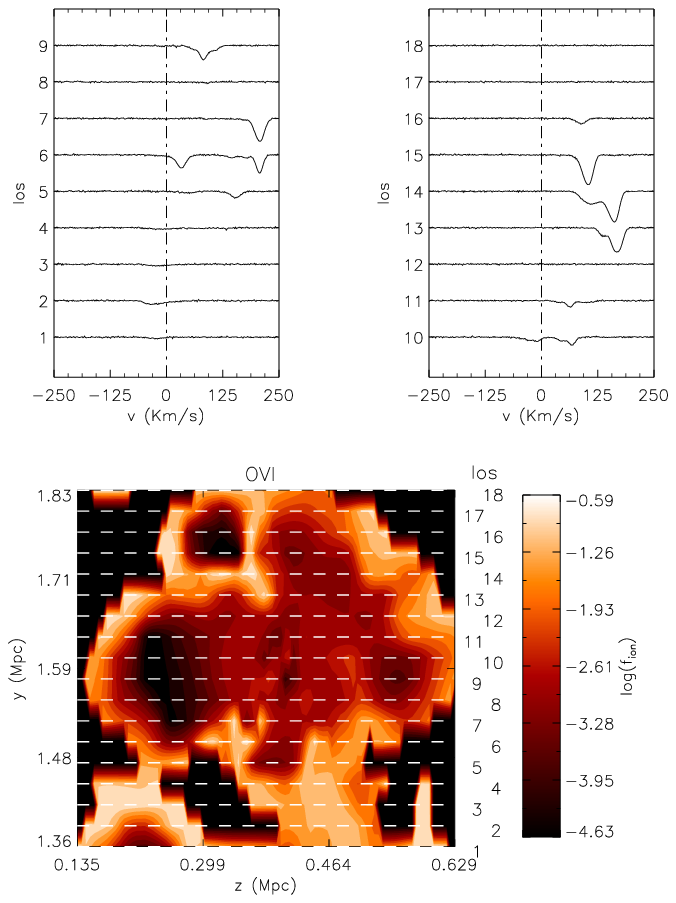}
\caption{} \label{spectra_bubble_OVI}
\end{figure*}
\clearpage
\begin{figure*}
\includegraphics[height=22cm]{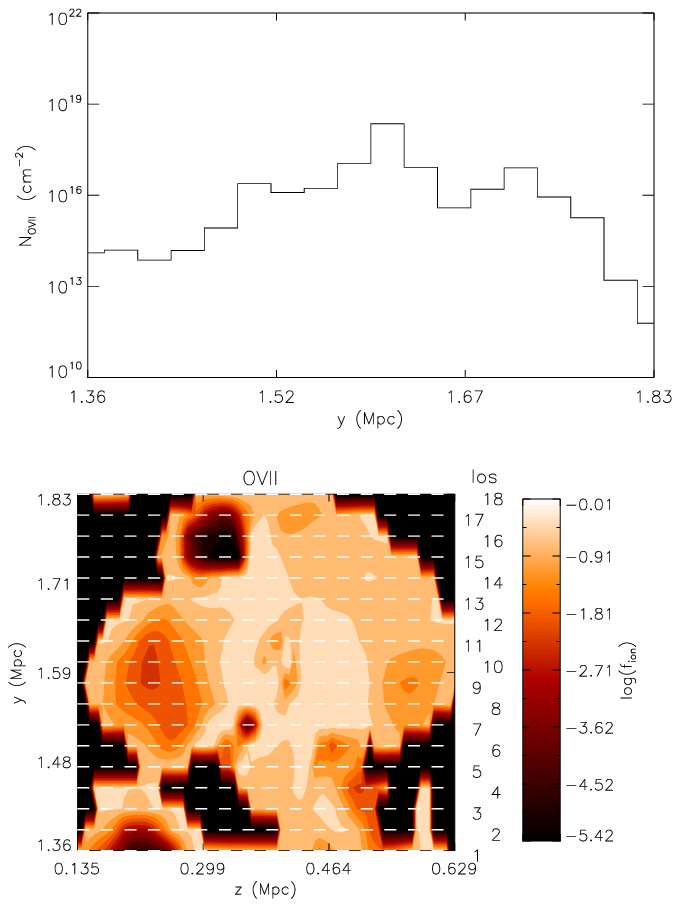}
\caption{} \label{spectra_bubble_OVII}
\end{figure*}
\clearpage
\begin{figure*}
\includegraphics[height=22cm]{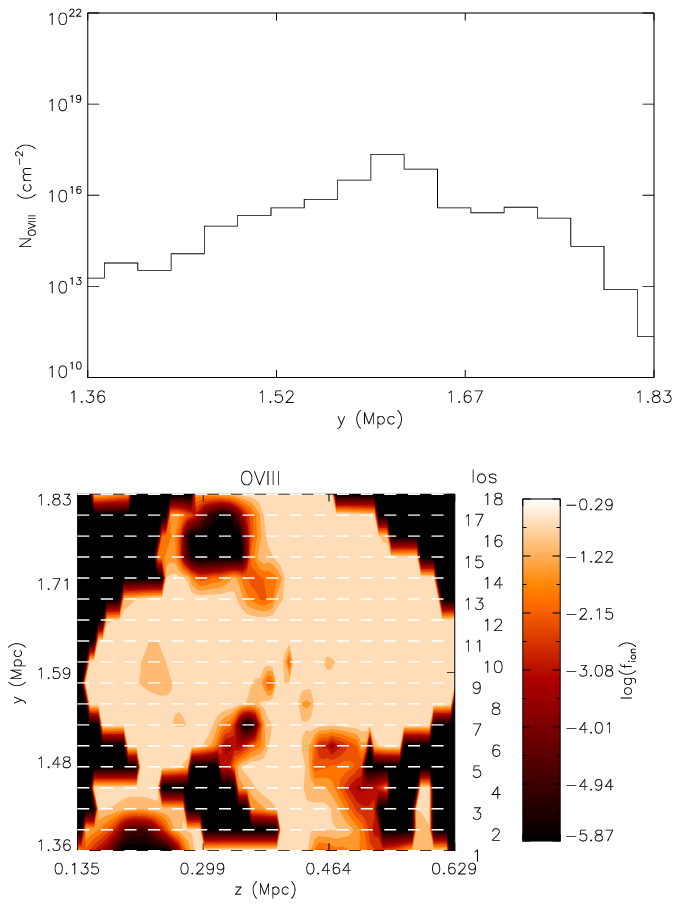}
\caption{} \label{spectra_bubble_OVIII}
\end{figure*}
\clearpage
\begin{figure*}
\includegraphics[height=22cm]{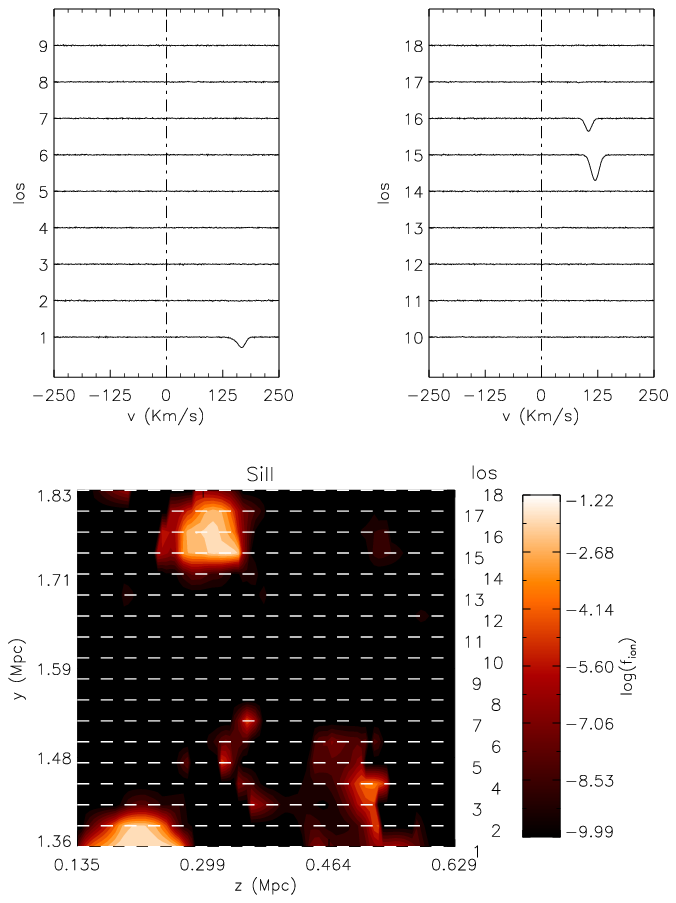}
\caption{} \label{spectra_bubble_SiII}
\end{figure*}
\clearpage
\begin{figure*}
\includegraphics[height=22cm]{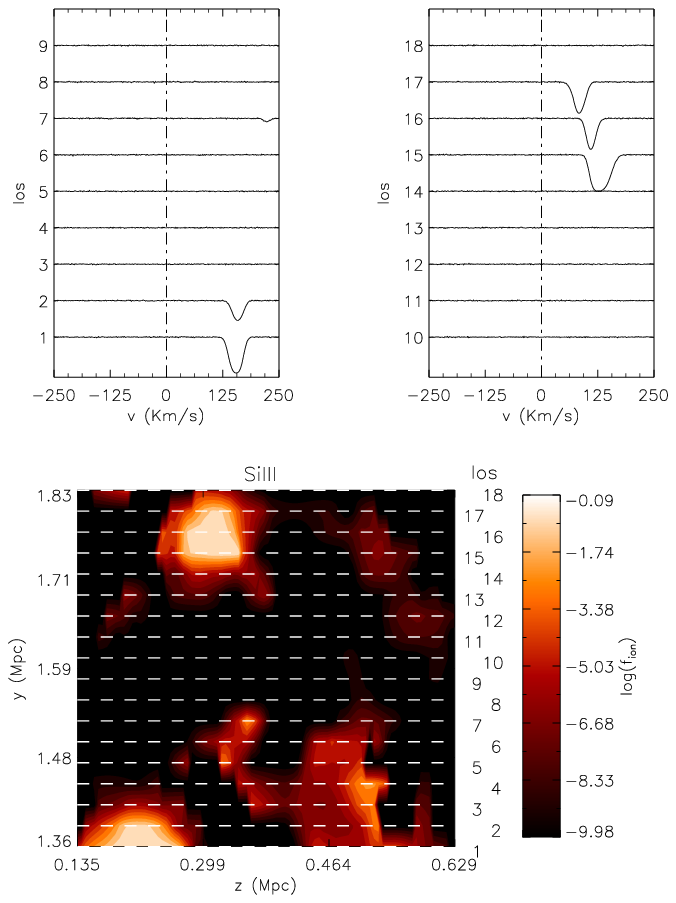}
\caption{} \label{spectra_bubble_SiIII}
\end{figure*}
\clearpage
\begin{figure*}
\includegraphics[height=22cm]{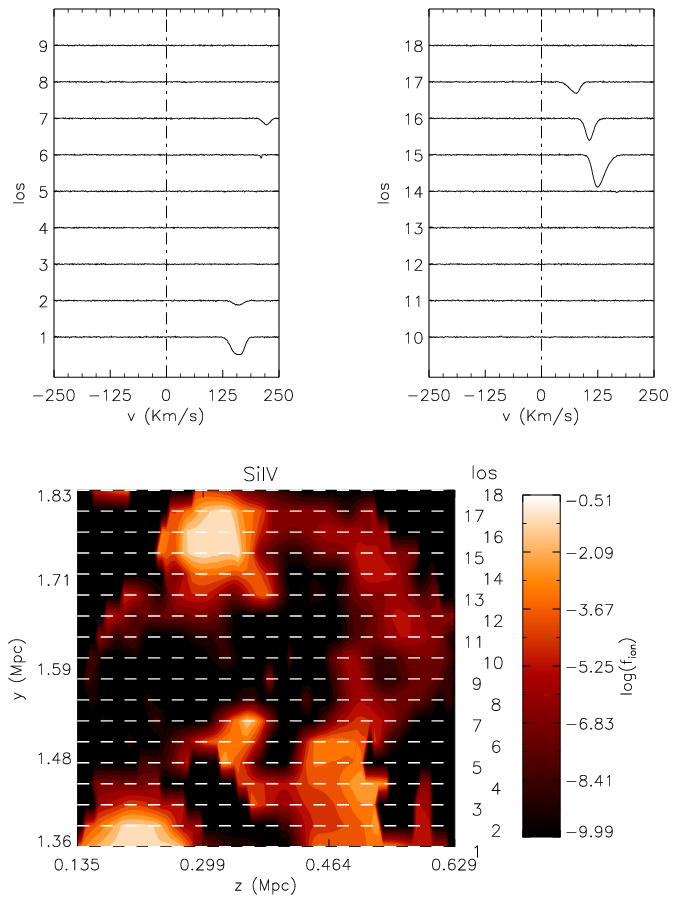}
\caption{}
\label{spectra_bubble_SiIV}
\end{figure*}
\clearpage

\end{document}